\documentclass[]{aa}  

\usepackage{graphicx}
\usepackage{epsfig}
\usepackage{multirow}
\usepackage{txfonts}
\begin{document}

   \title{Spitzer Space Telescope observations of bilobate comet 8P/Tuttle}

   \author{O. Groussin,		\inst{1}
        P. L. Lamy,     	\inst{2}
        M. S. P. Kelley,        \inst{3}
        I. Toth,        	\inst{1,4}
        L. Jorda,       	\inst{1}
        Y. R. Fern\'andez, 	\inst{5}
        \and
        H. A. Weaver      	\inst{6}
        }

   \institute{Aix Marseille Univ, CNRS, CNES, LAM, Marseille, France\\
     \email{olivier.groussin@lam.fr}
     \and Laboratoire Atmosphères, Milieux et Observations Spatiales, CNRS \& UVSQ, 11 bvd d’Alembert, 78280 Guyancourt, France     
     \and Department of Astronomy, University of Maryland, College Park, MD 20742-2421, USA
     \and MTA CSFK Konkoly Observatory, H1121 Budapest, Konkoly Thege M. ut 15-17., Hungary
     \and Department of Physics and Florida Space Institute, University of Central Florida, Orlando, FL 32816, USA
     \and The Johns Hopkins University Applied Physics Laboratory, Laurel, MD 20723, USA}

   \date{Received ---; accepted ---}

\titlerunning{Spitzer Space Telescope observations of comet 8P/Tuttle}
\authorrunning{Groussin et al.}


\abstract
{Comet 8P/Tuttle is a Nearly Isotropic Comet (NIC), whose physical properties are poorly known and could be different from those of Ecliptic Comets (EC) owing to their different origin. Two independent observations have shown that 8P/Tuttle has a bilobate nucleus.}
{Our goal is to determine the physical properties of the nucleus (size, shape, thermal inertia, albedo) and coma (water and dust) of 8P/Tuttle.}
{We observed the inner coma of 8P/Tuttle with the infrared spectrograph (IRS) and the infrared camera (MIPS) of the Spitzer Space Telescope (SST). We obtained one spectrum (5~--~40~$\mu$m) on 2 November 2007 and a set of 19 images at 24~$\mu$m on 22~--~23 June 2008 sampling the nucleus rotational period. The data were interpreted using thermal models for the nucleus and the dust coma, and considering two possible shape models of the nucleus derived from respectively Hubble Space Telescope visible and Arecibo radar observations.}
{We favor a nucleus shape model composed of two contact spheres with respective radii of $2.7\pm0.1$~km and $1.1\pm0.1$~km and a pole orientation with RA~$=285\pm12$$^{\circ}$ and DEC~$=+20\pm5$$^{\circ}$. The nucleus has a thermal inertia lying in the range 0~--~100~J~K$^{-1}$~m$^{-2}$~s$^{-1/2}$ and a $R$-band geometric albedo of $0.042\pm0.008$. The water production rate amounts to 1.1$\pm$0.2$\times$10$^{28}$~molecules~s$^{-1}$ at 1.6~AU from the Sun pre-perihelion, which corresponds to an active fraction of $\approx$9~\%. At the same distance, the $\epsilon f \rho$ quantity amounts to $310\pm34$~cm at 1.6~AU, and reaches $325\pm36$~cm at 2.2~AU post-perihelion. The dust grain temperature is estimated to $258\pm10$~K, which is 37~K larger than the thermal equilibrium temperature at 1.6 AU. This indicates that the dust grains contributing to the thermal infrared flux have a typical size of $\approx$10~$\mu$m. The dust spectrum exhibits broad emissions around 10~$\mu$m (1.5-$\sigma$ confidence level) and 18~$\mu$m (5-$\sigma$ confidence level) that we attribute to amorphous pyroxene.}
{}

 
\maketitle

%
%

\section{Introduction}

Comet 8P/Tuttle belongs to the family of nearly isotropic comets (NIC), following the classification of \citet{Levison1997}, and more precisely the family of Halley-type comets (HTC) \citep{Levison1994}. Compared with ecliptic comets (EC), little is known about the nucleus properties of NIC. Whereas we have information on those of more than 200 EC \citep{Lamy2004,Fernandez2013}, it is only the case for less than 30 NIC \citep{Lamy2004}. Owing to their different dynamical reservoirs, the Oort cloud for NIC \citep{Levison2001} and the Kuiper belt for EC \citep{Levison1991}, the question naturally arises as to whether these two populations have intrinsically different physical properties. 

Prior to its last passage on January 2008 at only 0.25~AU from the Earth, the nucleus of comet 8P/Tuttle was thought to be very large. Actually, \citet{Licandro2000} derived a radius of 7.3 km from visible photometry at an heliocentric distance ($r_h$) of 6.3~AU, assuming a typical visible geometric albedo of 0.04, making 8P/Tuttle potentially one of the largest NIC after Hale-Bopp (37~km) and 109P/Swift-Tuttle (13~km) \citep{Lamy2004}. Visible observations performed in 2006 by \citet{Weissman2008}, when the comet was at $r_h=$~5.0~AU, also supported a large radius of 6.0~km. However, the radar observations performed by \citet{Harmon2010} in early January 2008 showed a very different picture, revealing a bilobate shape. The two lobes were found elongated, with semi-axes of 2.1~$\times$~2.1~$\times$~2.9~km for the larger lobe and 1.6~$\times$~1.6~$\times$~2.1~km for the smaller lobe. This implies a much smaller nucleus than originally found. Visible observations performed by \citet{Lamy2008a} with the Hubble Space Telescope (HST) on 10~--~11 December 2007 during 12 HST visits extending over a 28~hour time interval also indicated a small nucleus with a radius of 3.0~km. \citet{Harmon2010} and \citet{Lamy2008a} derived a rotation period of 11.4~hr. From millimeter observations with the Plateau de Bure interferometer, \citet{Boissier2011} obtained an upper limit for the nucleus thermal inertia of 10~J~K$^{-1}$~m$^{-2}$~s$^{-1/2}$.

The water production rate of comet 8P/Tuttle was measured close to perihelion ($r_h=$~1.03~AU on 27 January 2008) by several observers. \citet{Biver2008} derived a water production rate of 4.0$\times$10$^{28}$~molecules~s$^{-1}$ from millimeter observations (IRAM) between 29 December 2007 and 2 January 2008, when 8P/Tuttle was at $r_h=$~1.10~--~1.12~AU. \citet{Barber2009} observed 8P/Tuttle in the near infrared with the United Kingdom Infrared Telescope (UKIRT) on 3 January 2008 ($r_h=$~1.09~AU) and derived a water production rate of 1.4$\pm$0.1$\times$10$^{28}$~molecules~s$^{-1}$. \citet{Lovell2008} derived a water production rate of 1.8$\times$10$^{28}$~molecules~s$^{-1}$ from radio observations (Arecibo and Green Bank) on 15 January 2008, at $r_h=$~1.04~AU. \citet{Lippi2008} derived a water production rate of 5.4~--~6.0$\times$10$^{28}$~molecules~s$^{-1}$ from near infrared observations using CRIRES at the ESO VLT on 27 January 2008, at $r_h=$~1.03~AU. With the same instrument but a few days later (28 January 2008 -- 4 February 2008), \citet{Bockelee2008} derived a water production rate of 3.9~--~4.4$\times$10$^{28}$~molecules~s$^{-1}$, at $r_h=$~1.03~AU.

To summarize, the water production rate of comet 8P/Tuttle close to perihelion lies in the range 1.4~--~6.0$\times$10$^{28}$~molecules~s$^{-1}$. For a radius of 3.0~km, it corresponds to a surface active area in the range 3~--~15~\%, derived from the water production rate of a spherical nucleus made of water ice only, located at perihelion, and assuming a temperature distribution similar to that of the Standard Thermal Model with a beaming factor of 1 \citep{Lebofsky1986}. For other gas species, not relevant to this paper, the reader is directed to \citet{AHearn1995}, \citet{Bohnhardt2008}, \citet{Bonev2008}, \citet{Jehin2009}, and \citet{Kobayashi2010}.

Two determinations of the $Af\rho$ quantity of comet 8P/Tuttle are available: 110~cm by \citet{AHearn1995} when the comet was close to perihelion in August 1994, and 32~cm by \citet{Schleicher2007} during the interval 3~--~5 December 2007 at $r_h=$~1.3~AU. These values are low compared with other comets \citep{AHearn1995}, which likely indicates a paucity of sub-micron size dust particles.

The peculiar bilobate nature of 8P/Tuttle was relatively unique when it was discovered in 2008, the only other questionable examples at that time being 1P/Halley based on its ``central depression'' \citep{Keller1987} and 19P/Borrelly \citep{Soderblom2002}. Since that, two other cometary nuclei have been confirmed to be bilobate: 103P/Hartley 2 \citep{AHearn2011} and 67P/Churyumov-Gerasimenko \citep{Sierks2015}. Overall, four of the six comets for which we have spatially resolved images of their nucleus have a bilobate shape, which seems therefore a common shape among cometary nuclei.

The aim of this paper is to present the results of our Spitzer Space Telescope (SST) observations of comet 8P/Tuttle performed on 2 November 2007 with the IRS instrument and during 22~--~23 June 2008 with the MIPS instrument, in order to determine the physical properties of its nucleus (size, shape, thermal inertia, albedo) and the activity level of its coma (water and dust). In particular, using thermal infrared observations, we estimate the size of the nucleus, independently of its geometric albedo.

%
%
\section{Observations with the Spitzer Space Telescope}

\subsection{IRS and MIPS observations}

The orbital elements of 8P/Tuttle are given in Table~\ref{orbital_elements}. There were only two visibility windows of about 3 months each to observe comet 8P/Tuttle with the SST during cycle 4 (June 2007 -- June 2008) because of the restriction on solar elongation (80~--~120$^{\circ}$). The first window, from 4 October 2007 to 24 January 2008, covered the pre-perihelion phase from $r_h=$~1.9 to 1.03~AU, with an increasing phase angle from 32 to 75$^{\circ}$. The second window, from 4 April 2008 to 30 June 2008, covered the post-perihelion phase from $r_h=$~1.5 to 2.3~AU, with a decreasing phase angle from 40 to 22$^{\circ}$. At the time of proposal preparation, the best size estimate of the nucleus radius was 7.3~km \citep{Licandro2000}, so that we anticipated a very large flux. In fact, this flux was expected to saturate the MIPS 24~$\mu$m detector during almost the entire first window and the IRS detector close to perihelion. As a consequence, the scheduled window was carefully selected to maximize the signal-to-noise on the nucleus without saturating the MIPS and IRS detectors. Because of further additional constraints on the phase angle, we ultimately decided on the following observing strategy: (i) perform the IRS observations on 2 November 2007, before the expected flux reached the saturation limit and (ii) perform the MIPS observations on 22~--~23 June 2008, at low phase angle after the expected flux dropped below the saturation limit. Table \ref{table_obs} summarizes the IRS and MIPS observations.

\begin{table}
\caption[]{Comet 8P/Tuttle orbital elements from the JPL's Horizons website$^1$ for the 27 January 2008 perihelion passage: perihelion distance ($q$), aphelion distance ($Q$), eccentricity ($e$), inclination ($i$), orbital period ($P$) and Tisserand's parameter with respect to Jupiter ($T_{\rm J}$).}
\label{orbital_elements}
\begin{center}
\begin{tabular}{cccccc}
\hline
\noalign{\smallskip}
$q$ & $Q$ & $e$ & $i$ & $P$ & $T_{\rm J}$\\
$[$AU$]$& $[$AU$]$ &  & $[^{\circ}]$ & $[$year$]$ &\\
\noalign{\smallskip}
\hline
\noalign{\smallskip}
1.03 &10.4 &0.82 &55 &13.6 &1.6 \\
\noalign{\smallskip}
\hline
\end{tabular}
\end{center}
\end{table}

At the time of the IRS observations, 8P/Tuttle was at $r_h=$~1.61~AU, a distance from the SST of 1.32~AU, and a solar phase angle of 39$^{\circ}$. We used IRS in the low resolution mode ($R = \lambda/\Delta\lambda \approx$ 64~--~128) that covers the wavelength range 5.2~--~38.0~$\mu$m in four long-slit segments: the short wavelength 2$^{nd}$ order (SL2, from 5.2 to 8.5~$\mu$m), the short wavelength 1$^{st}$ order (SL1, from 7.4 to 14.2~$\mu$m), the long wavelength 2$^{nd}$ order (LL2, from 14.0 to 21.5~$\mu$m), and the long wavelength 1$^{st}$ order (LL1, from 19.5 to 38.0~$\mu$m). We acquired three spectra with an integration time of 18.9~sec for each segment, i.e. a total integration time of 56.7~sec per segment. The pointing of the target was performed using the ephemeris derived from the JPL's Horizons website\footnote{https://ssd.jpl.nasa.gov/horizons.cgi}. We could not use the peak-up cameras at the time of observation because of saturation issues. However, since the SST pointing error is only $\approx$1\arcsec (smaller than the slit width) and since the ephemeris were even more accurate, the peak-up cameras were unnecessary. The same sequence was repeated two days later, on 4 November 2007, at the same RA and DEC as the original observations, to obtain shadow observations in order to properly subtract the sky background.


At the time of the MIPS observations, 8P/Tuttle was at $r_h=$~2.24~AU, a distance from the SST of 1.58~AU, and a solar phase angle of 23$^{\circ}$. We used the MIPS imaging capabilities at 24~$\mu$m and 70~$\mu$m to take observations centered on the nucleus. At 24~$\mu$m, we performed 20 observations, with a common integration time of 48.2~sec. Each observation consists of 14 dithered frames mosaicked together (Section~2.3). The MIPS 24~$\mu$m detector works at an effective wavelength of 23.7~$\mu$m with a pixel scale of 2.55\arcsec/pixel. At 70~$\mu$m, we performed 4 observations centered on the nucleus, with a common integration time of 37.7~sec. Each observation consists of 12 dithered frames mosaicked together (Section~2.3). The MIPS 70~$\mu$m detector works at an effective wavelength of 71.0~$\mu$m with a pixel scale of 9.96\arcsec/pixel. At the time of proposal preparation, the rotation period of the nucleus was unknown. To minimize the amount of observing time requested and still maintain a reasonable chance of obtaining the light curve extrema, the 20 observations at 24~$\mu$m were distributed unevenly over 15~h and separated by either 0.5 or 1.0~h. The 4 observations at 70~$\mu$m were likewise distributed unevenly over 15~h and inserted between the 24~$\mu$m observations. The same sequence was repeated one day later, on 24 June 2008, at the same RA and DEC as the original observations, to secure shadow observations. 

Detail about the SST can be found in \citet{Werner2004} and in the Spitzer observer's manual\footnote{http://ssc.spitzer.caltech.edu/documents/som/}. More information on the instruments can be found in \citet{Houck2004} for IRS and \citet{Rieke2004} for MIPS.

\begin{table*}
\caption[]{IRS and MIPS observations. For each observation, we list the instrument, the wavelength of observation ($\lambda$), the starting date of the observation, the heliocentric distance ($r_h$), the distance to SST ($\Delta$), the phase angle ($\alpha$), the nucleus infrared flux and the dust $\epsilon f \rho$ quantity (see text for detail).}
\label{table_obs}
\begin{center}
\begin{tabular}{cccccccc}
\hline
\noalign{\smallskip}
Instrument  &$\lambda$      &Date   &$r_h$	&$\Delta$	&$\alpha$	&Nucleus flux	&$\epsilon f \rho$\\
            &$[$$\mu$m$]$         &$[$UT$]$     &$[$AU$]$       &$[$AU$]$             &$[^{\circ}]$            &$[$mJy$]$  &$[$cm$]$    \\
\noalign{\smallskip}
\hline
\noalign{\smallskip}
IRS	&5.2~--~38.0	&2007 11 02.76	&1.606	&1.322	&39.4	&N/A	        &$310\pm34$\\
\noalign{\smallskip}
MIPS 	&23.7		&2008 06 22.49	&2.243	&1.579	&23.5	&--	        &$326 \pm 36$\\
MIPS	&23.7		&2008 06 22.52	&2.243	&1.579	&23.5	&$110\pm6$	&$329 \pm 36$\\
MIPS	&23.7		&2008 06 22.53	&2.243	&1.579	&23.5	&$100\pm5$	&$327 \pm 36$\\
MIPS	&23.7		&2008 06 22.58	&2.244	&1.580	&23.5	&$85\pm4$	&$327 \pm 36$\\
MIPS	&23.7		&2008 06 22.62	&2.244	&1.580	&23.5	&$105\pm5$	&$326 \pm 36$\\
MIPS	&23.7		&2008 06 22.64	&2.244	&1.580	&23.5	&$108\pm5$	&$328 \pm 36$\\
MIPS	&23.7		&2008 06 22.66	&2.244	&1.580	&23.5	&$110\pm6$	&$326 \pm 36$\\
MIPS	&23.7		&2008 06 22.70	&2.245	&1.580	&23.4	&$122\pm6$	&$328 \pm 36$\\
MIPS	&23.7		&2008 06 22.74	&2.245	&1.581	&23.4	&$130\pm7$	&$328 \pm 36$\\
MIPS	&23.7		&2008 06 22.79	&2.246	&1.581	&23.4	&$120\pm6$	&$327 \pm 36$\\
MIPS	&23.7		&2008 06 22.83	&2.246	&1.581	&23.4	&$110\pm6$	&$324 \pm 36$\\
MIPS	&23.7		&2008 06 22.85	&2.246	&1.582	&23.4	&$105\pm5$	&$326 \pm 36$\\
MIPS	&23.7		&2008 06 22.87	&2.247	&1.582	&23.4	&$115\pm6$	&$326 \pm 36$\\
MIPS	&23.7		&2008 06 22.92	&2.247	&1.582	&23.4	&$125\pm6$	&$325 \pm 36$\\
MIPS	&23.7		&2008 06 22.95	&2.247	&1.582	&23.4	&$113\pm6$	&$327 \pm 36$\\
MIPS	&23.7		&2008 06 22.97	&2.248	&1.582	&23.4	&$110\pm6$	&$325 \pm 36$\\
MIPS	&23.7		&2008 06 22.99	&2.248	&1.582	&23.4	&$108\pm5$	&$324 \pm 36$\\
MIPS	&23.7		&2008 06 23.03	&2.248	&1.583	&23.4	&$104\pm5$	&$322 \pm 36$\\
MIPS	&23.7		&2008 06 23.08	&2.249	&1.583	&23.4	&$100\pm5$	&$322 \pm 36$\\
MIPS	&23.7		&2008 06 23.12	&2.249	&1.583	&23.4	&$109\pm5$	&$322 \pm 36$\\
\noalign{\smallskip}
MIPS 	&71.0		&2008 06 22.49	&2.243	&1.579	&23.5	&--	&--\\
MIPS	&71.0		&2008 06 22.62	&2.244	&1.580	&23.5	&--	&--\\
MIPS	&71.0		&2008 06 22.83	&2.246	&1.581	&23.4	&--	&--\\
MIPS	&71.0		&2008 06 22.95	&2.247	&1.582	&23.4	&--	&--\\
\noalign{\smallskip}
\hline
\end{tabular}
\end{center}
\end{table*}

\subsection{IRS data reduction}

Spectra of comet 8P/Tuttle acquired with the IRS instrument were initially processed and calibrated with the Spitzer Science Center's IRS pipeline (version S17.0.4).  We subtracted a shadow observation from each target observation to remove the sky background.  Some residual background flux still remained in the 2D spectral images. Therefore, a second subtraction was performed using median-combined sky frames taken contemporaneously with the comet. This removes any zodiacal light or instrument background not fully accounted for in the shadow observations. Bad pixels were identified and replaced via nearest-neighbor interpolation, or ignored altogether.

We extracted spectra from the 2D images using the Spitzer Science Center's SPICE software \footnote{SPICE is available at http://ssc.spitzer.caltech.edu/}.  Our synthetic apertures, centered on the peak of the source, used the default point source aperture widths, which vary with $\lambda/\lambda_0$: 4.0 pixels at $\lambda_0=6.0$~$\mu$m (SL2) and 8.0 pixels at $\lambda_0=12.0$~$\mu$m (SL1) for the short-low extractions, and 4.3 pixels at $\lambda_0=16.0$~$\mu$m (LL2) and 7.2 pixels at $\lambda_0=27.0$~$\mu$m (LL1) for long-low. Finally, to improve the signal-to-noise ratio (SNR), we computed the median spectrum of our three spectra. 

As the goal of our observations is to measure the spectrum of the nucleus, we did not apply any extended source calibrations to the extracted data in order to preserve the spectral shape of the emission from the unresolved nucleus. As we will see in Section~5.1, the nucleus contributes to $\approx$50\% of the total signal in the SL mode and $\approx$25\% of the total signal in the LL mode.

\subsection{MIPS data reduction}

\begin{figure*} [!ht] 
\begin{center}
  \includegraphics[width=13.0cm]{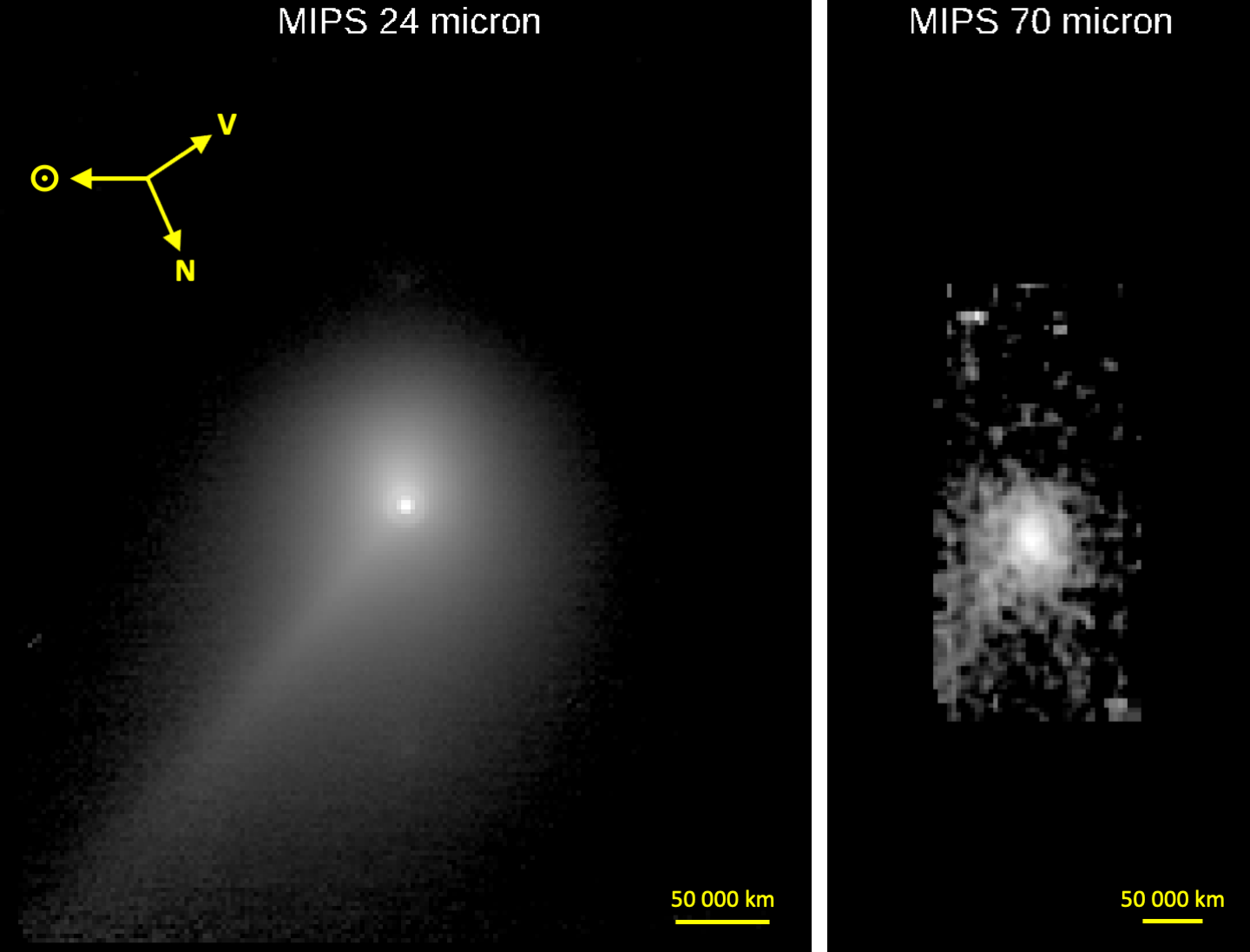}
\end{center}
\caption{MIPS infrared calibrated images of comet 8P/Tuttle taken on 22 June 2008 (24~$\mu$m on the left, 70~$\mu$m on the right). The images are displayed with a logarithmic stretch. The 24~$\mu$m image has a size of 176$\times$194~pixels with a projected pixel size of about 2450~km. The yellow arrows indicate the direction to the Sun ($\sun$), celestial North (N), and velocity vector (V). The 70~$\mu$m image has a size of 46$\times$95~pixels with a projected pixel size of about 3840~km. The coma, dust tail and first Airy ring are visible in the 24~$\mu$m image. The 70~$\mu$m image is extremely noisy and inappropriate for nucleus extraction (see text for detail).}
\label{fig_8P}
\end{figure*}

The images acquired with MIPS were processed with the Spitzer Science Center's pipeline (version S18.0.2), producing individual basic calibrated data (BCD) frames. We subtracted a shadow observation from each target observation to remove the sky background. The BCD images were subsequently mosaicked in the rest frame of the comet with the MOPEX software (version 16.3.7). Bad pixels, that is those permanently damaged, or affected by cosmic rays, were ignored during the mosaicking step. Figure~\ref{fig_8P} shows an example of a calibrated image at 24~$\mu$m and 70~$\mu$m.

\begin{figure} [!ht] 
\begin{center}
\includegraphics[width=8.5cm]{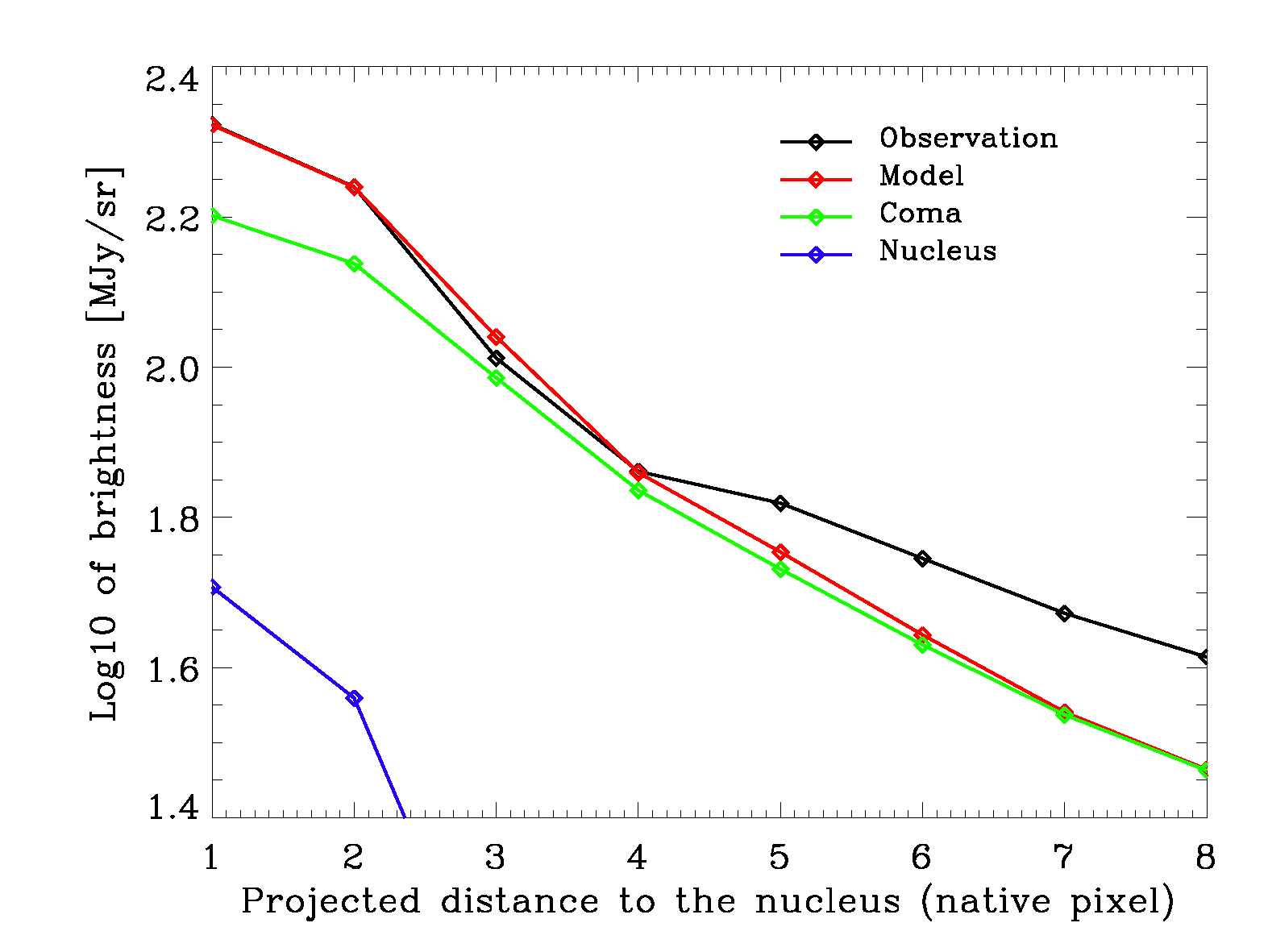}
\end{center}
\caption{Azimuthal average profile of the central coma of comet 8P/Tuttle, observed by the MIPS instrument at 24~$\mu$m, and compared with a photometric model comprising a separate contribution from the nucleus and the coma. In the central 2 pixels, used to derive the nucleus contribution, the fit is excellent and one cannot distinguish between the model and the observation.}
\label{fig_azimaver}
\end{figure}

The coma, dust tail and first Airy ring are visible in the 24~$\mu$m image, indicating a high nucleus to coma ratio in the central pixel. The extraction of the nucleus signal was performed using our standard method of fitting a parametric model of the expected surface brightness to the observed images, as implemented for instance to MIPS observation of comet 67P/Churyumov-Gerasimenko by \citet{Lamy2008b}. A 2-dimensional array of brightness was constructed, which superimposes an unresolved nucleus represented by a Dirac function and a simple coma model that follows the canonical 1/$\rho$ radial variation ($\rho$ is the projected distance from the nucleus), both convolved by the point spread function (PSF) of the telescope. The PSFs were generated with the STINYTIM\footnote{STINYTIM is available at https://irsa.ipac.caltech.edu/data/ SPITZER/docs/dataanalysistools/tools/contributed/general/stinytim/} tool, following the procedure described in \citet{Lamy2008b}. The model images were generated on a finer grid than the original MIPS pixel, with a resampling factor of 10. The fit to the real images was performed after integrating the model over $10 \times 10$ sub-pixels to recover the original pixel of the MIPS. The fits to the observations were performed on azimuthally averaged radial profiles excluding the sector affected by the tail and led to the determination of the respective scaling factors of the nucleus and coma models. As shown in one example given in Fig.~\ref{fig_azimaver}, the fits were satisfactory up to a distance $\rho$ = 4 pixels (9800~km at 1.58~AU) with residuals of typically 1\%. Note that the total signal in the central pixel is dominated by the coma, the nucleus contribution amounting to a fraction of only $\approx$25\% of the total signal. This is however sufficient for a robust nucleus extraction as justified by \cite{Hui2018}. The derived fluxes for the nucleus are given in Table \ref{table_obs} for each image. The typical 1-$\sigma$ error is 5~\%. The nucleus extraction was not possible on the first image because the fit of the model to the average radial profile was very poor and the resulting nucleus flux abnormally low ($<$50~mJy) compared with the values for the other 19 MIPS images.

Concerning the 70~$\mu$m images, their SNR of $\approx$6 was too low to extract the nucleus. Its contribution in the central pixel was estimated to $\approx$10\% of the total flux on the basis of the 24~$\mu$m image, using geometrical considerations. This results from the larger field of view at 70~$\mu$m, which translates in a larger contribution from the coma whereas that of the nucleus remains the same. In comparison, the 24~$\mu$m images have a SNR of $\approx$200.

%
%
\section{Nucleus and coma thermal models}

The IRS spectrum is a combination of the thermal flux coming from the nucleus and the dust and gas coma. Thermal models of the nucleus and the coma are therefore required to estimate their respective contribution to the total Spectral Energy Distribution (SED). For MIPS, only a thermal model for the nucleus is required, since its flux can be extracted from the overall signal (nucleus + coma) as explained in Section~2.3.

\subsection{Nucleus thermal model}

\subsubsection{The nucleus shape model}

\begin{figure*} [!ht]
\includegraphics[width=18.4cm]{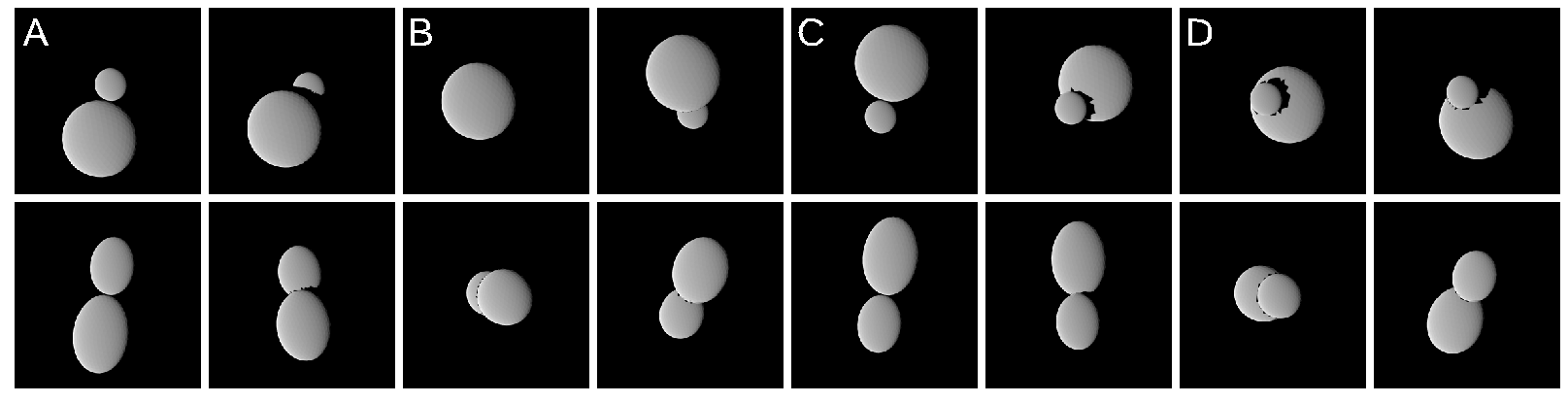}
\caption{HST shape model (top row) and radar shape model (bottom row) as viewed by SST on 22~--~23 June 2008. Letters A, B, C and D correspond to different time (extrema) during the rotation and are connected to Fig.~\ref{fig_lc2}.}
\label{fig_shape}
\end{figure*}

The nucleus thermal model first requires a shape model. For 8P/Tuttle, we cannot make the usual assumption of a spherical nucleus since there are evidences for a bilobate shape. Currently, two different shape models for the nucleus of comet 8P/Tuttle are available. The first one (hereafter HST shape model) is derived from the inversion of a visible light curve obtained by the Hubble Space Telescope \citep{Lamy2008a}, whereas the second one (hereafter radar shape model) is derived from radar observations \citep{Harmon2010}. The two shape models correspond to a contact binary, but they differ in the shape and size of the primary and secondary (Table~\ref{table_shape} and Fig.~\ref{fig_shape}). 

The HST shape model consists of two spheres in contact with a ratio of 2.3 between their radii. The absolute scale of this model is not constrained since it depends on the albedo. For a typical geometric albedo of 0.04 ($R$ band), the radius of the two spheres amounts to 2.8~km and 1.2~km, respectively. The pole orientation for the HST shape model defined by RA~$=285\pm12$$^{\circ}$ and DEC~$=+20\pm5$$^{\circ}$ yields an aspect angle (defined as the angle between the spin vector and the comet-SST vector) of 92$^{\circ}$ on 2 November 2007 (IRS) and 65$^{\circ}$ on 22~--~23 June 2008 (MIPS).

The radar shape model consists of two prolate ellipsoids in contact, aligned along their long axis. The semi-axes are $a=$2.06~km, $b=$2.06~km and $c=$2.88~km for the primary, and $a=$1.64~km, $b=$1.64~km and $c=$2.13~km for the secondary. There is a 10\% uncertainty on these values. The pole axis is perpendicular to the long axis. The radar observations constrain the pole orientation to lie within a cone corresponding to a projection angle of 55$\pm$7$^{\circ}$ from the observer. However, we calculated that, inside this cone, the solution that best fits the HST light curve corresponds to RA~$=268\pm5$$^{\circ}$ and DEC~$=-16\pm2$$^{\circ}$, which gives an aspect angle of 60$^{\circ}$ on 2 November 2007 (IRS) and 104$^{\circ}$ on 22~--~23 June 2008 (MIPS).

There is a separation of 40$^{\circ}$ between the pole directions given by the two models. However, both have a rotational period of 11.4~h.

\begin{table*}
  \caption[]{The HST and radar shape models.}
  \label{table_shape}
  \begin{center}
\begin{tabular}{ll}
\hline
\noalign{\smallskip}
\multicolumn{2}{l}{{\bf HST shape model} \citep{Lamy2008a}} \\
\noalign{\smallskip}
- Shape & Two spheres in contact\\
- Largest sphere (radius)  & 2.8~km \\
- Smallest sphere (radius)  & 1.2~km \\
- Pole orientation & RA=285$\pm$12$^{\circ}$ and DEC=+20$\pm$5$^{\circ}$\\
\noalign{\smallskip}
\hline
\noalign{\smallskip}
\multicolumn{2}{l}{{\bf Radar shape model} \citep{Harmon2010}} \\
\noalign{\smallskip}
- Shape & Two prolate spheroids in contact, aligned along their long axis\\
- Largest spheroid (semi-axis)& 2.06~$\times$~2.06~$\times$~2.88~km\\
- Smallest spheroid (semi-axis)  & 1.64~$\times$~1.64~$\times$~2.13~km \\
- Pole axis   & Perpendicular to the long axis\\
- Pole orientation  & RA=268$\pm$5$^{\circ}$ and DEC=-16$\pm$2$^{\circ}$ {\it (this work)}\\
\noalign{\smallskip}
\hline
\end{tabular}
\end{center}
\end{table*}

\subsubsection{Thermal model}

We implemented our nucleus thermal model already extensively described in several past articles \citep[e.g.,][]{Groussin2004,Lamy2008c}, so that we presently limit ourselves to a short description.

We consider the two above shape models with their respective pole orientation.
For each one, the surface of the nucleus is divided into 2560 facets, and for each facet, we solve for the surface energy balance between the flux received from the Sun, the re-radiated flux, and the heat conduction into the nucleus.
As we will show, the active fraction of 8P/Tuttle is restricted to $\approx$9~\% of its surface (Section~5.5). Likewise the case of comet 9P/Tempel~1 which has an active fraction of 9\% \citep{Lisse2005} the sublimation of water ice can be neglected in the energy balance for the calculation of the thermal flux emitted from the nucleus surface \citep{Groussin2007}.
As the nucleus rotates around its spin axis, the illumination changes and the heat conduction equation is computed for each facet considering a one-dimensional time-dependent equation.
The projected shadows are taken into account.
We used a time step of $\approx$12~sec, which is small enough compared with the rotation period ($\approx$11.4~hr) to ensure relaxation of the numerical solution in a few tens of rotations (depending on the thermal inertia). 
As a result, we obtained the temperature $T_{i}$ of each facet as a function of time, over one period of rotation. 

From this surface temperature distribution, we calculated the infrared flux received by the observer from each facet as a function of time and wavelength, 5~--~40~$\mu$m for IRS and 24~$\mu$m for MIPS. The total flux $F_{\rm nucl}$ is then the sum of all individual fluxes of each facet of the shape model (Eq.~\ref{eq_flux_nucl}).
When the thermal inertia is not null and since the phase angle is not negligible for the IRS (39$^{\circ}$) and MIPS (24$^{\circ}$) observations, the infrared flux depends on the solution adopted for the rotation, i.e., prograde (P) or retrograde (R). Owing to the lack of information on this point, both cases are studied.

\begin{equation}
F_{\rm nucl}(\lambda) = \gamma \sum_{i=1}^{n} \epsilon B(\lambda,T_{i}) d\Omega_{i}
\label{eq_flux_nucl}
\end{equation}

\subsubsection{Parameters of the nucleus thermal model}

Our model has six free parameters: the infrared emissivity $\epsilon$, the phase integral $q$, the scaling coefficient for the nucleus flux $\gamma$ (it corresponds to a scaling coefficient$\sqrt{\gamma}$ for the shape model), the geometric albedo $p_{\rm  v}$ , the beaming factor $\eta$ and the thermal inertia $I$. 
Among these six parameters, we consider that three of them $\epsilon$, $q$ and $p_{\rm  v}$ can be reasonably assumed whereas the other three $\gamma$, $\eta$ and $I$ must be constrained by the observations.

We adopted a value of 0.95 for the thermal emissivity $\epsilon$, which is the mid-point of the values typically quoted in the literature (0.9~--~1.0). As the value is near 1.0, it has a negligible influence on the calculated thermal flux.

The phase integral $q$ measures the angular dependence of the scattered  radiation. We chose $q$=0.27, the value found for 19P/Borrelly by \citet{Buratti2004}. We adopted a geometric albedo $p_{\rm  v}$=0.04, a typical value for cometary nuclei \citep{Lamy2004}. The choice of $q$ and $p_{\rm  v}$ has a negligible influence on the size determination, as long as the product $p_{\rm  v} q$ remains in the range 0.0~--~0.1, which is the case for all cometary nuclei.

The beaming factor $\eta$ follows the strict definition given by \citet{Lagerros1998} and therefore, only reflects the influence of surface roughness that produces an anisotropic thermal emission. Theoretically, $\eta$ ranges from 0 to 1, but in practice, it is larger than 0.7 to avoid unrealistic roughness \citep{Lagerros1998}.
It differs from the factor $\eta$ used in the Standard Thermal Model \citep[STM;][]{Lebofsky1989} or in the Near-Earth Asteroid Thermal Model \citep[NEATM;][]{Harris1998}  where $\eta$ is a combination of roughness {\it and} thermal inertia and thus can be larger than one.   
In this study, we considered four values for $\eta$: 0.7, 0.8, 0.9 and 1.0. 

We considered several values for the thermal inertia $I$=0, 50, 100, 200, 400, 800~J~K$^{-1}$~m$^{-2}$~s$^{-1/2}$, covering more than the range 0~--~350~J~K$^{-1}$~m$^{-2}$~s$^{-1/2}$ found for comets \citep[see the review paper of][]{Groussin2019}.

The parameter $\gamma$ scales the nucleus flux to match the data, and directly scales the size of the nucleus set by the shape model.
As a consequence, it can be independently determined for each combination of $\eta$ and $I$.

\subsection{Dust coma thermal model}

In addition to the above nucleus thermal model, the interpretation of the IRS spectra requires a thermal model for the dust coma, in order to estimate the SED. There are a multitude of possibilities for the dust SED based on many choices of grain size distribution, grain composition, and grain structure. Moreover, our IRS spectrum (as we show) has a limited amount of compositional diagnostics in it that could help us to independently constrain the dust grain properties. So, to make the problem tractable, we adopted a simple graybody model, as given by Eq.~(\ref{eq_flux}):
\begin{equation}
F_{\rm coma}(\lambda) = \mathcal{A} \frac{\epsilon_{\rm dust}}{\Delta^2} B(\lambda,T_{\rm dust}) \Bigg( \frac{\lambda}{\lambda_{\rm 0}} \Bigg) ^{p}
\label{eq_flux}
\end{equation}
where $F_{\rm coma}(\lambda)$ is the thermal flux (Jy) at the wavelength $\lambda$ ($\mu$m), $\mathcal{A}$ is the dust cross-section in the field of view (m$^2$), $\epsilon_{\rm dust}$ is the dust emissivity (we assume 0.95), $\Delta$ is the observer-comet distance (m) and $B(\lambda,T_{\rm dust})$ is the Planck function at temperature $T_{\rm dust}$ (K). The two unknowns are $\mathcal{A}$ and $T_{\rm dust}$.

The factor $(\lambda / \lambda_{\rm 0})^{p}$ is required for the coma aperture correction, which was not taken into account in the data reduction (Section~2.2). This correction is purely geometrical. We used $\lambda_{\rm 0} = 5.0$~$\mu$m for the SL mode and $\lambda_{\rm 0} = 14.0$~$\mu$m for the LL mode. The power $p$ is used to convert the rectangular aperture of the IRS slit into an equivalent circular aperture, assuming a canonical $1/\rho$ radial brightness profile for the coma. For the SL and LL modes, we computed $p=0.37$. It means, for example, that without the coma aperture correction the coma flux is 46\% larger at 14.0~$\mu$m than at 5.0~$\mu$m in the SL mode, or 40\% larger at 35.0~$\mu$m than at 14.0~$\mu$m in the LL mode.

%
%
\section{MIPS data analysis}

\subsection{Adjusting the model to the data}

\begin{figure} [!ht] 
\includegraphics[width=\linewidth]{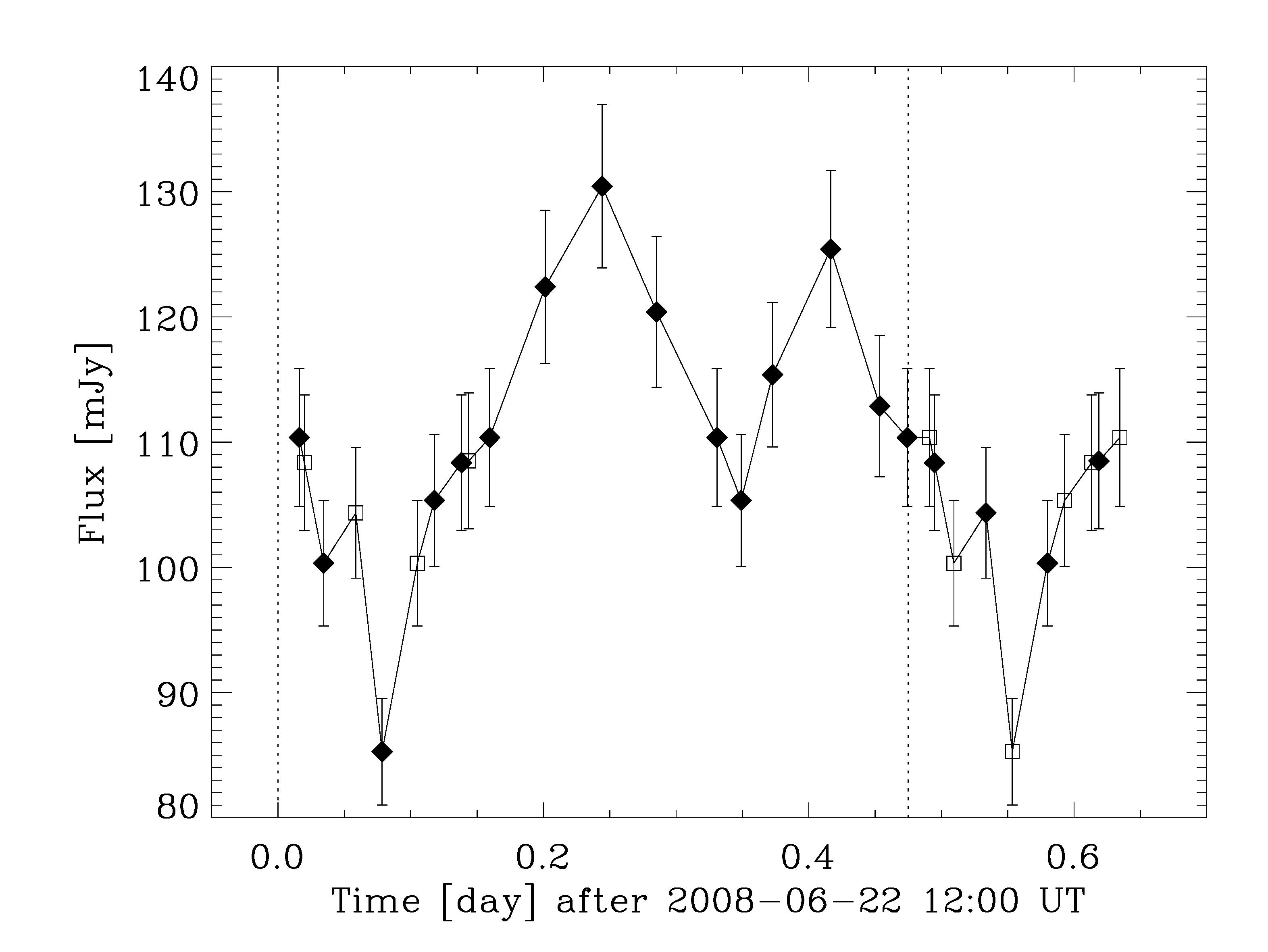}
\caption{MIPS thermal light curve at 24~$\mu$m. Filled diamond symbols correspond to the MIPS nucleus fluxes of Table~\ref{table_obs}. Square symbols have been phase-folded using a rotation period of 11.4~hr (marked by a vertical dashed line at 0~day and 0.475~day). The light curve is double peaked with one minimum (at 0.08~day) being stronger than the other (at 0.35~day).}
\label{fig_lc1}
\end{figure}

The thermal infrared light curve of the nucleus of comet 8P/Tuttle derived from the MIPS observations (Table~\ref{table_obs}) is plotted in Fig.~\ref{fig_lc1}. It has been phase-folded using a rotation period of 11.4~hr. The light curve is double peaked with one minimum being deeper than the other, in agreement with a contact binary and similar to the HST visible light curve \citep{Lamy2008a}.

\begin{figure} [!ht] 
\includegraphics[width=\linewidth]{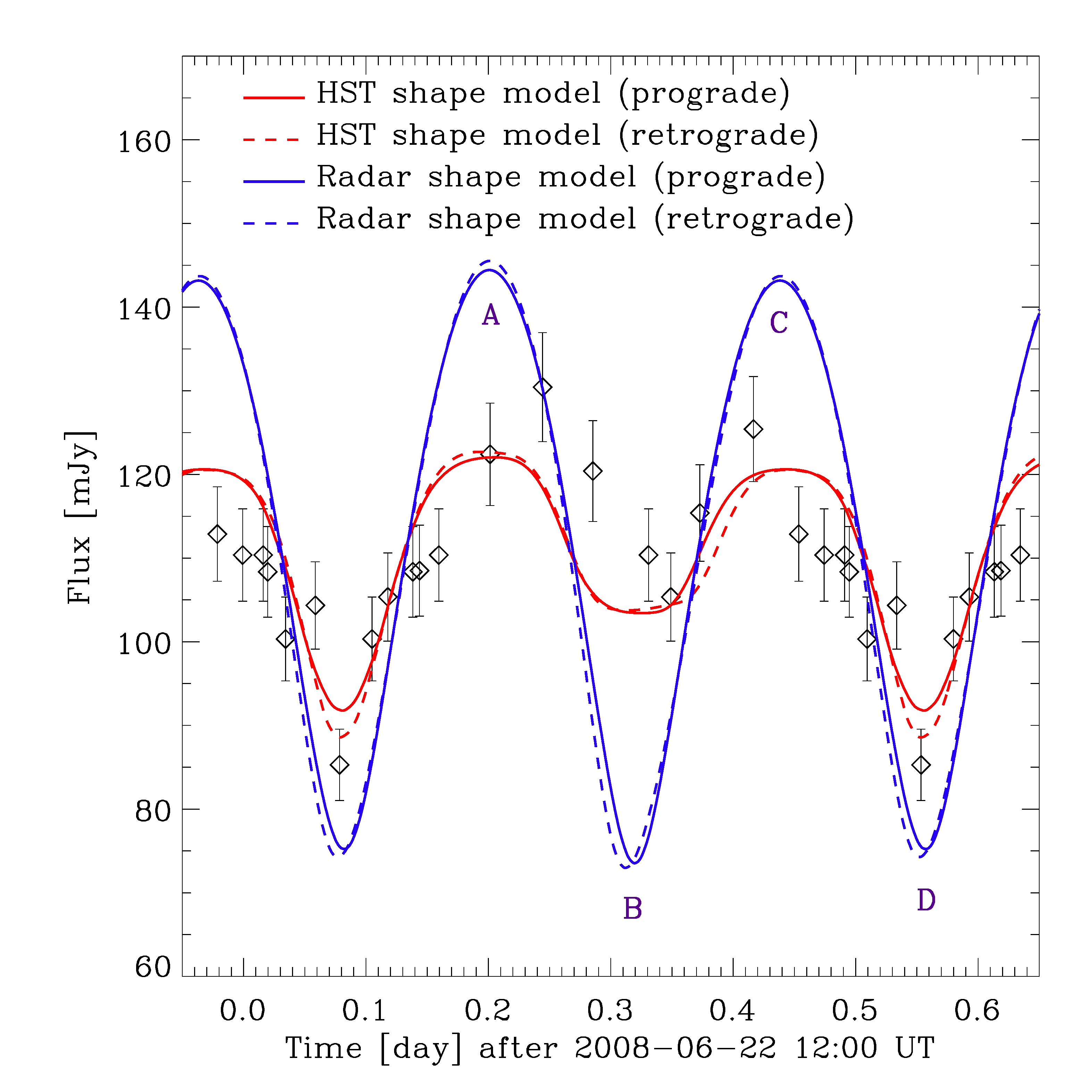}
\caption{Synthetic thermal light curves corresponding to the HST and radar shape models (prograde and retrograde) for the combination $\eta$=0.7 and $I$=50~J~K$^{-1}$~m$^{-2}$~s$^{-1/2}$. The model thermal light curves and the MIPS data have been extended beyond one rotation period for clarity. Letters A, B, C and D correspond to the extrema and are connected to Fig.~\ref{fig_shape}.}
\label{fig_lc2}
\end{figure}

As explained above, for each shape model (HST and radar) we calculated a synthetic thermal light curve for different values of $\eta$ and $I$. For each combination of $\eta$ and $I$, we adjusted the synthetic light curve to the data by tuning the scaling factor $\gamma$ and the phase. The best fit is determined by minimizing the chi-square value. Results are given in Table~\ref{table_mips} and illustrated in Fig.~\ref{fig_lc2}. The uncertainty on $\gamma$ is $\approx$5~\%.

\begin{table*}
  \caption[]{Results for the fits of the synthetic thermal light curve to the MIPS data using the HST and radar shape models. $\eta$ is the beaming factor. $I$ is the nucleus thermal inertia (J~K$^{-1}$~m$^{-2}$~s$^{-1/2}$). $\gamma$ is the derived scaling factor for the nucleus shape model to match the MIPS infrared flux. $\chi^2$ is the chi-square value; for reference, our model has 16 degrees of freedom (19 data points and 3 free parameters).}
\label{table_mips}
\begin{center}
\begin{tabular}{cccccccccc}
\hline
\noalign{\smallskip}
$\eta$	&$I$	&\multicolumn{4}{c}{HST shape model}	&\multicolumn{4}{c}{Radar shape model} \\
	&	&\multicolumn{2}{c}{Prograde}	&\multicolumn{2}{c}{Retrograde}	&\multicolumn{2}{c}{Prograde}	&\multicolumn{2}{c}{Retrograde} \\
	&	&$\gamma$	&$\chi^2$	&$\gamma$	&$\chi^2$	&$\gamma$	&$\chi^2$	&$\gamma$	&$\chi^2$	\\	
\noalign{\smallskip}
\hline
\noalign{\smallskip}
   0.7  &     0  &  0.83  &  37.0  &  0.83  &  37.6  &  0.91  & 193.2  &  0.90  & 190.7   \\
   0.7  &    50  &  0.89  &  31.6  &  0.97  &  35.1  &  0.97  & 171.6  &  1.05  & 187.8   \\
   0.7  &   100  &  0.97  &  27.8  &  1.07  &  30.6  &  1.04  & 155.1  &  1.14  & 161.7   \\
   0.7  &   200  &  1.09  &  21.3  &  1.21  &  26.5  &  1.15  & 132.1  &  1.26  & 125.4   \\
   0.7  &   400  &  1.23  &  19.4  &  1.33  &  26.4  &  1.28  & 114.6  &  1.37  & 108.5   \\
   0.7  &   800  &  1.35  &  25.1  &  1.41  &  31.3  &  1.37  & 110.8  &  1.44  & 112.6   \\
\noalign{\smallskip}
   0.8  &     0  &  0.90  &  36.9  &  0.90  &  37.6  &  0.99  & 194.1  &  0.99  & 191.6   \\
   0.8  &    50  &  0.97  &  31.8  &  1.06  &  35.5  &  1.06  & 173.1  &  1.14  & 190.0   \\
   0.8  &   100  &  1.05  &  28.1  &  1.17  &  30.8  &  1.14  & 156.5  &  1.25  & 164.8   \\
   0.8  &   200  &  1.19  &  21.4  &  1.33  &  26.7  &  1.26  & 133.1  &  1.38  & 127.1   \\
   0.8  &   400  &  1.36  &  19.4  &  1.47  &  26.7  &  1.40  & 115.1  &  1.50  & 108.8   \\
   0.8  &   800  &  1.49  &  25.6  &  1.56  &  31.9  &  1.51  & 110.6  &  1.58  & 112.3   \\
\noalign{\smallskip}
   0.9  &     0  &  0.97  &  37.1  &  0.97  &  37.8  &  1.07  & 194.9  &  1.06  & 192.4   \\
   0.9  &    50  &  1.05  &  31.9  &  1.14  &  35.9  &  1.14  & 174.4  &  1.23  & 192.0   \\
   0.9  &   100  &  1.14  &  28.2  &  1.27  &  31.1  &  1.23  & 157.9  &  1.35  & 167.4   \\
   0.9  &   200  &  1.29  &  21.5  &  1.44  &  26.9  &  1.36  & 134.3  &  1.49  & 128.8   \\
   0.9  &   400  &  1.48  &  19.5  &  1.60  &  27.1  &  1.52  & 115.7  &  1.63  & 109.3   \\
   0.9  &   800  &  1.63  &  26.2  &  1.71  &  32.4  &  1.64  & 110.4  &  1.72  & 112.2   \\
\noalign{\smallskip}
   1.0  &     0  &  1.04  &  37.2  &  1.05  &  37.8  &  1.14  & 195.7  &  1.14  & 193.0   \\
   1.0  &    50  &  1.13  &  32.2  &  1.23  &  36.1  &  1.23  & 175.7  &  1.32  & 193.7   \\
   1.0  &   100  &  1.22  &  28.5  &  1.37  &  31.5  &  1.32  & 159.2  &  1.45  & 169.9   \\
   1.0  &   200  &  1.39  &  21.6  &  1.55  &  27.1  &  1.46  & 135.3  &  1.61  & 130.5   \\
   1.0  &   400  &  1.59  &  19.5  &  1.73  &  27.3  &  1.64  & 116.2  &  1.76  & 109.8   \\
   1.0  &   800  &  1.76  &  26.6  &  1.86  &  32.9  &  1.77  & 110.3  &  1.87  & 112.1   \\
\noalign{\smallskip}
\hline
\end{tabular}
\end{center}
\end{table*}

\subsection{The HST and radar shape models}

The HST shape model provides a better fit to the data than the radar shape model, whatever the combination of $\eta$, $I$, prograde or retrograde. The minimum chi-square value for the radar shape model ($\chi^2$=108.5 for $\eta$=0.7, $I$=400~J~K$^{-1}$~m$^{-2}$~s$^{-1/2}$, retrograde rotation) is $\approx$3 times larger than the maximum chi-square value for the HST shape model ($\chi^2$=37.2 for $\eta$=1.0, $I$=0~J~K$^{-1}$~m$^{-2}$~s$^{-1/2}$, prograde rotation), and in most cases the chi-square values differ by a factor $>$5 between the two shape models.

For the radar shape model, the 10\% uncertainty on the nucleus size \citep{Harmon2010} translates to a 20\% uncertainty on the flux and therefore restricts $\gamma$ to the range 0.8~--~1.2. In Table~\ref{table_mips}, all the solutions with $\gamma>1.2$ can thus be discarded for the radar shape model. On the contrary, the scale of the HST shape model is free, and all combinations $(\eta,I)$ are possible. Nevertheless, this scaling difference is not sufficient to explain the discrepancy between the two shape models. Indeed, even when $\gamma$ for the radar shape model is in the range 0.8~--~1.2 as in Fig.~\ref{fig_lc2} ($\eta$=0.7, $I$=50~J~K$^{-1}$~m$^{-2}$~s$^{-1/2}$), the fit is still worse than with the HST shape model.

The first discrepancy is the larger amplitude of the thermal light curve for the radar shape model compared with the HST shape model, which results from a larger semi-major axis ratio of 1.7 for the radar shape model compared with 1.4 for the HST shape model. The amplitude of the thermal light curve for the radar shape model is too large to properly fit the data.

The second discrepancy is minimum B, which is too deep for the radar shape model (Fig.~\ref{fig_lc2}). For the two shape models, minima B and D correspond to one lobe eclipsing the other (Fig.~\ref{fig_shape}). For the radar shape model, the eclipse is partial for the two minima and since the illuminated cross-sections are close, both minima are identical. For the HST shape model, minimum B corresponds to the larger body fully eclipsing the other and minimum D corresponds to the smaller body partially eclipsing the other. In this case, minimum D is more pronounced than B due to projected shadows close to the sub-solar region where the thermal flux mainly comes from.

\begin{figure} [!ht] 
\includegraphics[width=\linewidth]{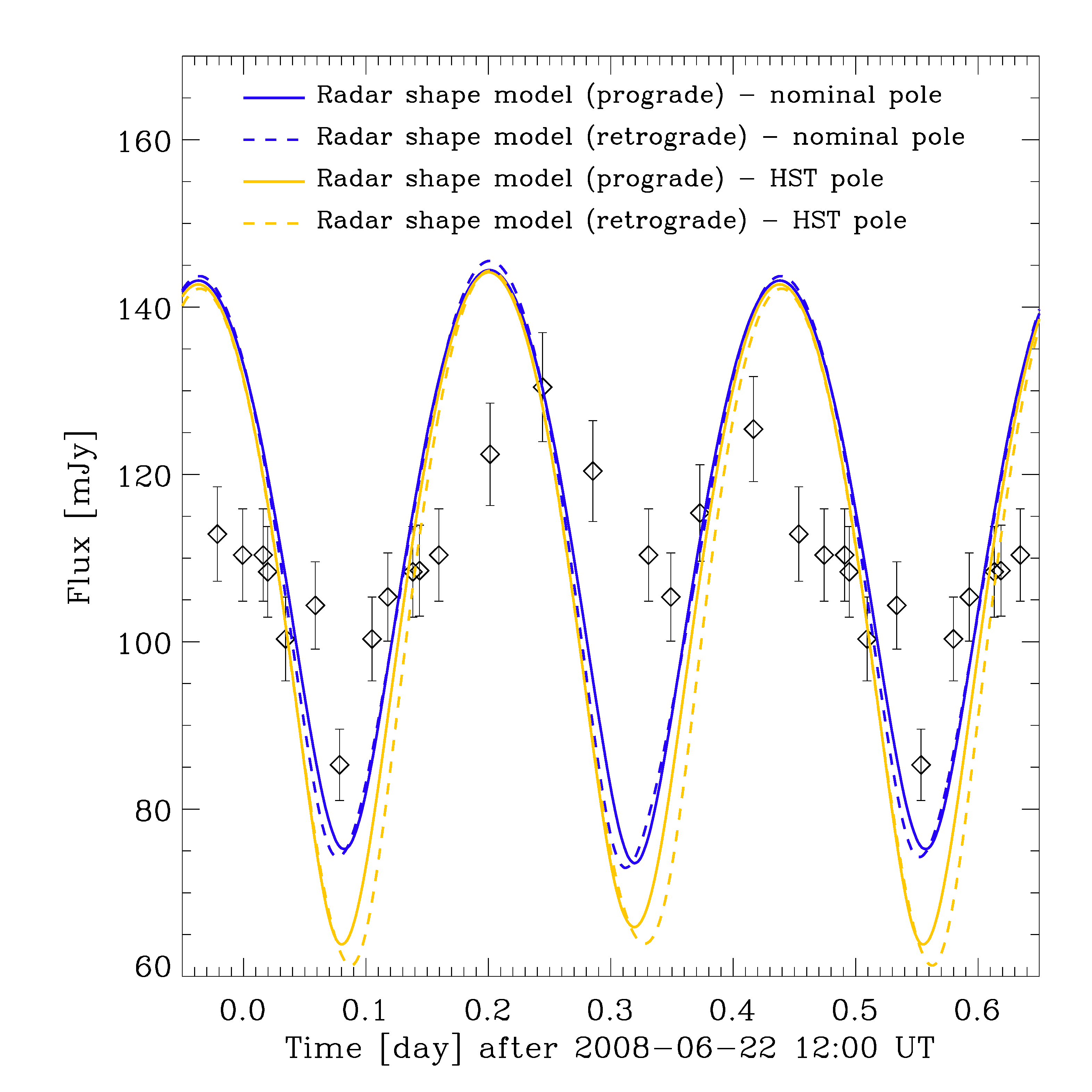}
\caption{Synthetic thermal light curves corresponding to the radar shape models (prograde and retrograde) for the combination $\eta$=0.7 and $I$=50~J~K$^{-1}$~m$^{-2}$~s$^{-1/2}$, and two different pole solutions (nominal radar pole and HST pole).}
\label{fig_lc2bis}
\end{figure}

The different pole orientation of the HST and radar shape models partially explains the above discrepancies, due to the differences in projected shadows, but it is however not sufficient. Indeed, as shown in Fig.~\ref{fig_lc2bis}, even the radar shape model with the pole orientation of the HST shape model does not fit the data. So, overall, the discrepancy between the HST and radar solutions is a combination of their different pole orientation {\it and} shape.

From Table~\ref{table_mips}, the minimum $\chi^2$ value of 19.4 is obtained with the HST shape model for $\eta$=0.7 or 0.8, $I$=400~J~K$^{-1}$~m$^{-2}$~s$^{-1/2}$, prograde rotation. Our model has 16 degrees of freedom (19 data points and 3 free parameters), which gives a reduced chi-square of 1.2, i.e. a reasonable value close to 1. We computed a $\Delta \chi^2$ of 36.2 for a confidence level of 99.7\% (i.e., 3-sigma for the normal distribution). At this confidence level, all the solutions with $\chi^2 > 55.6$ ($= 19.4+36.2$) can be rejected, which discards the radar shape model following strictly this statistical criteria.

To conclude, the radar shape model has two identical minima and a large amplitude, both in disagreement with the MIPS thermal light curve and the HST visible light curve \citep{Lamy2008a}. Moreover, the HST shape model always provides a better qualitative and quantitative fit to the data than the radar shape model, whatever the combination of $\eta$, $I$, prograde or retrograde. So, we favor the HST shape model over the radar shape model.

\begin{figure} [!ht] 
  \includegraphics[width=\linewidth]{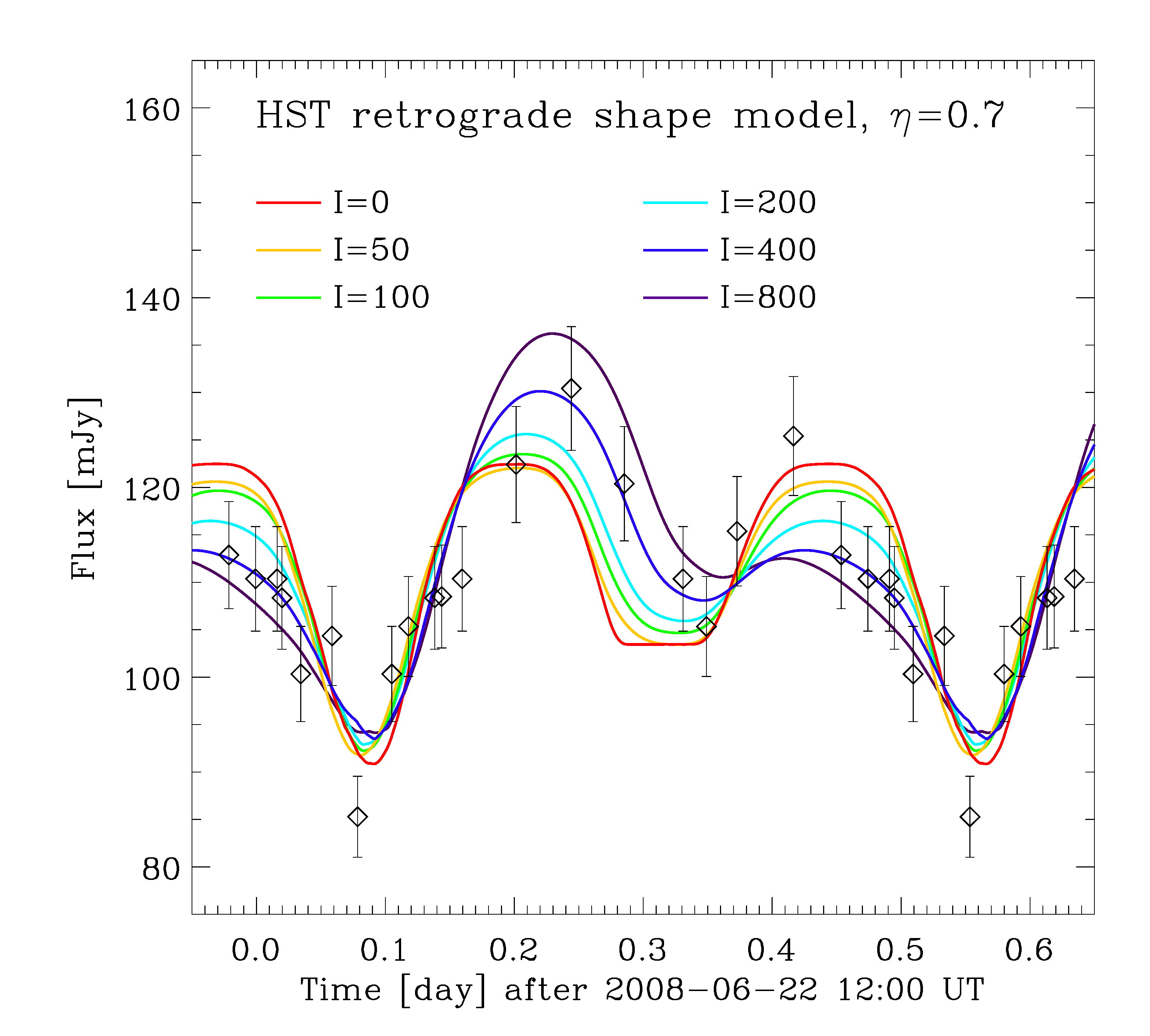}
  \includegraphics[width=\linewidth]{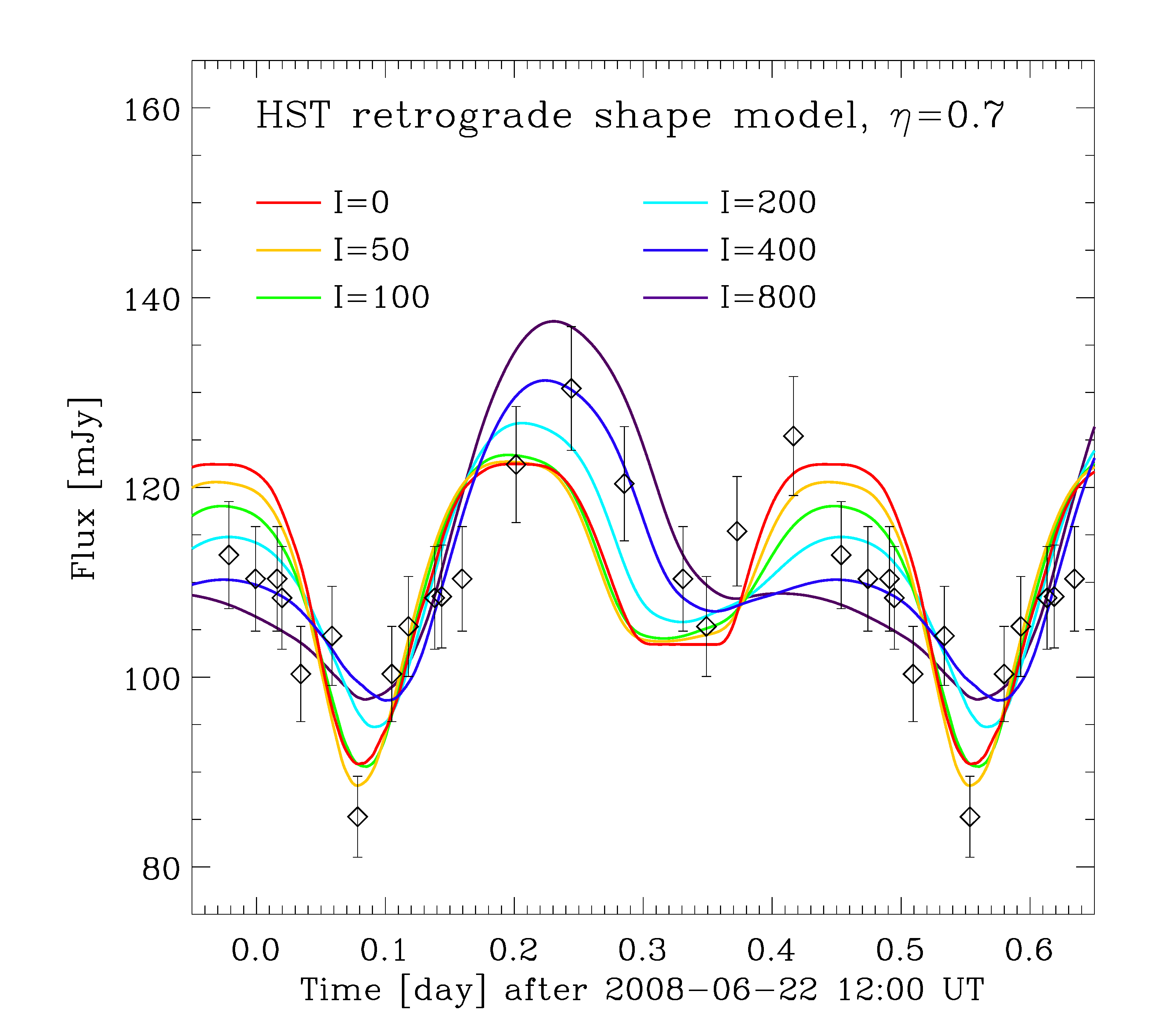}
\caption{Synthetic thermal light curves using the HST shape model (prograde on top, retrograde on bottom) for $\eta$=0.7 and different values of thermal inertia $I$ between 0 and 800~J~K$^{-1}$~m$^{-2}$~s$^{-1/2}$.}
\label{fig_lc3}
\end{figure}

From Table~\ref{table_mips}, for the HST model, the $\chi^2$ value decreases when thermal inertia increases, up to $I=400$~J~K$^{-1}$~m$^{-2}$~s$^{-1/2}$ that provides the best fit to the data. The improvement of a factor $\approx$2 of the chi-square value is however not significant enough to reject any values of $\eta$ or $I$ at this stage, in particular because extrema C and D are not well reproduced by the largest values of thermal inertia (Fig.~\ref{fig_lc3}); additional constraints coming from IRS spectra are required.


\subsection{Dust $\epsilon f \rho$ quantity}

We computed the quantity $\epsilon f \rho$ from all the MIPS images. This quantity, discussed in \citet{Kelley2013}, is used to estimate the dust production in the infrared wavelength range by analogy with the $A f \rho$ quantity defined by \citet{AHearn1984} at visible wavelengths. The $\epsilon f \rho$ and $A f \rho$ quantities are independent of the aperture size $\rho$ if the coma has a canonical $1/\rho$ radial brightness profile.

In our case, we computed the quantity $\epsilon f \rho$ with a dust temperature of 258~K derived from the IRS spectra (Section~5.4). We computed $\epsilon f \rho$ for different aperture sizes and found that its value varies by less than 5\% for apertures between 10 and 30 pixels, consistent with a $1/\rho$ coma radial brightness profile over this aperture range (Fig.~\ref{fig_azimaver}). The $\epsilon f \rho$ values are reported in Table~\ref{table_obs}.

The $\epsilon f \rho$ quantity is very stable over time and only varies between 322~cm and 329~cm during the 15~h of MIPS observations. The mean value is $\epsilon f \rho = 325\pm36$~cm. The uncertainty mainly comes from the uncertainty on the dust temperature (258$\pm$10~K). 

%
%
\section{IRS data analysis}

\subsection{Adjusting the model to the data}

The IRS spectrum was acquired on 2 November 2008 around 18:15~UT, in less than 10 minutes. Unfortunately, the nucleus rotation period is not known with a sufficient accuracy to re-phase the spectrum with the HST or MIPS light curves, taken respectively 38 and 233 days after. As a consequence, we adopted a conservative approach and assumed two extreme solutions for the cross-section viewed by SST at the time of observation, corresponding to the minimum and maximum cross-sections. As found in the previous section, only the HST shape model reasonably fits the MIPS light curve so we performed the analysis with this shape model.

\begin{figure} [!ht] 
\includegraphics[width=\linewidth]{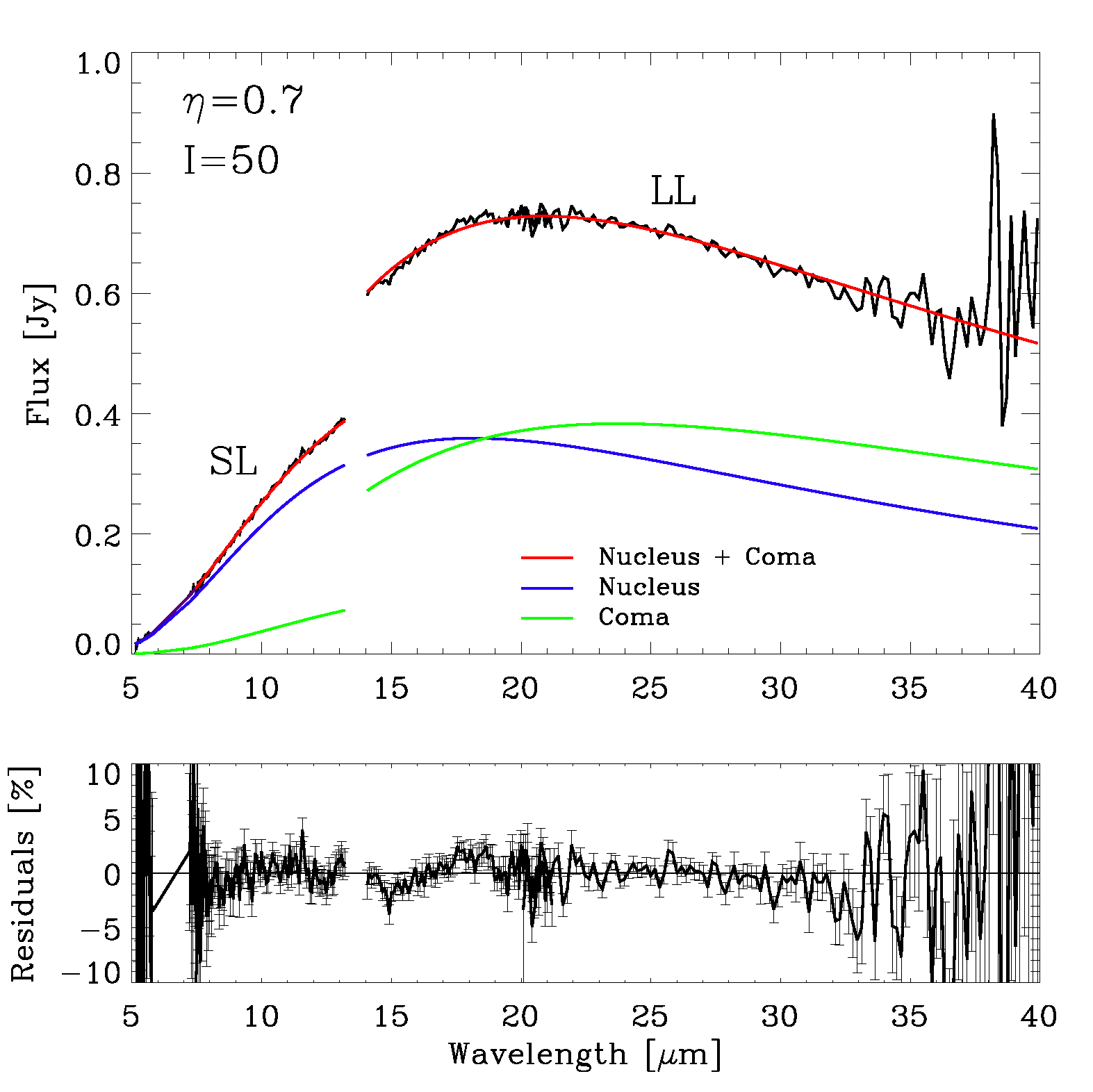}
\caption{IRS data and synthetic SED for the combination $\eta=0.7$ and $I=50$~J~K$^{-1}$~m$^{-2}$~s$^{-1/2}$ (minimum cross-section). The blue line is the contribution from the nucleus, the green line is that from the dust coma, and the red line is the sum of both (nucleus + coma). Residuals correspond to the difference between the model (nucleus + dust coma) and the data. The discontinuity between the SL and LL mode results from their different field of views, with less coma in SL mode (smaller field of view) than in LL mode (larger field of view).}
\label{fig_fit_irs}
\end{figure}

As explained in Section~3, there are five parameters to be constrained by the spectrum: $\gamma$, $\eta$ and $I$ for the nucleus and $\mathcal{A}$ and $T_{\rm dust}$ for the dust coma. There are more parameters than constraints and several solutions are possible. However, for a given combination of $\eta$ and $I$ that defines the shape of the nucleus SED, one can determine $T_{\rm dust}$ that controls the shape of the dust coma SED. The scaling factors $\gamma$ and $\mathcal{A}$ can then be adjusted to match the data. As a result, for each combination of $\eta$ in the range 0.7~--~1.0 and $I$ in the range 0~--~800~J~K$^{-1}$~m$^{-2}$~s$^{-1/2}$, we looked for the values of $T_{\rm dust}$, $\gamma$ and $\mathcal{A}$ that minimize the chi-square value between the synthetic SED ($F_{\rm nucl}+F_{\rm coma}$) and the data. Since the SL and LL modes have different slit widths (3.6~--~3.7\arcsec vs. 10.5~--~10.7\arcsec), there is one $\mathcal{A}$ value for each of them ($\mathcal{A}_{\rm SL}$ and $\mathcal{A}_{\rm LL}$). We estimated the uncertainty to 10~\% on $\gamma$ (IRS), 5~K on $T_{\rm dust}$, 3~$\times$~10$^6$~m$^2$ on $\mathcal{A}_{\rm SL}$, and 5~$\times$~10$^6$~m$^2$ on $\mathcal{A}_{\rm LL}$, in order to keep the residuals within the 1 sigma error bars. Results are given in Table~\ref{table_irs} and illustrated in Fig.~\ref{fig_fit_irs}. 

\begin{table*}
  \caption[]{Results for the fit of the synthetic SED to the IRS data using the HST shape model. Combinations ($\eta, I$) in bold are compatible with both IRS and MIPS observations. {\it Rotation} defines the sense of rotation (P for prograde, R for retrograde, N/A for a null thermal inertia). {\it Cross-section} is the cross-section facing SST at the time of observation: minimum or maximum (see text for detail). $\gamma$~(IRS) is the derived scaling factor for the IRS infrared flux. $T_{dust}$ is the derived temperature of the dust. $\mathcal{A}_{\rm SL}$ is the derived dust cross-section in the SL field of view. $\mathcal{A}_{\rm LL}$ is the derived dust cross-section in the LL field of view. $\gamma$~(MIPS) is the scaling factor for the MIPS infrared flux, from Table~\ref{table_mips}.}
\label{table_irs}
\begin{center}
\begin{tabular}{ccccccccc}
\hline
\noalign{\smallskip}
$\eta$	&$I$    &Rotation	&Cross-section	&$\gamma$~(IRS)	&$T_{\rm dust}$	&$\mathcal{A}_{\rm SL}$	&$\mathcal{A}_{\rm LL}$	&$\gamma$~(MIPS) \\
        &$[$J~K$^{-1}$~m$^{-2}$~s$^{-1/2}$$]$ & & & &$[$K$]$ & $[$10$^6$~m$^2$$]$ & $[$10$^6$~m$^2$$]$ & \\
\noalign{\smallskip}
\hline
\noalign{\smallskip}
\multirow{2}{*}{\bf 0.7}
&\multirow{2}{*}{\bf 0}
&\multirow{2}{*}{\bf N/A}
  &min    &1.31	&262	&24	&75	&\multirow{2}{*}{0.83}	\\
&&&max    &0.64	&267	&29	&79	\\
\hline
\noalign{\smallskip}
\multirow{4}{*}{\bf 0.7}
&\multirow{4}{*}{\bf 50}
&\multirow{2}{*}{\bf Prograde}
&min	&1.61	&254	&13	&64	&\multirow{2}{*}{0.89}	\\
&&&max	&0.75	&264	&25	&75	&	\\
&&\multirow{2}{*}{\bf Retrograde}
&min	&2.25	&267	&14	&57	&\multirow{2}{*}{0.97}	\\
&&&max	&0.95	&265	&24	&73	&	\\
\hline
\noalign{\smallskip}
\multirow{4}{*}{\bf 0.7}
&\multirow{4}{*}{\bf 100}
&\multirow{2}{*}{\bf Prograde}
&min	&1.79	&248	&7	&60	&\multirow{2}{*}{0.97}	\\
&&&max	&0.90	&258	&21	&73	&	\\
&&\multirow{2}{*}{Retrograde}
&min	&3.42	&270	&0	&34	&\multirow{2}{*}{1.07}	\\
&&&max	&1.37	&262	&16	&63	&	\\
\hline
\noalign{\smallskip}
\multirow{4}{*}{0.7}
&\multirow{4}{*}{200}
&\multirow{2}{*}{Prograde}
&min	&2.12	&242	&0	&53	&\multirow{2}{*}{1.09}	\\
&&&max	&1.21	&249	&12	&67	&	\\
&&\multirow{2}{*}{Retrograde}
&min	&0.47	&302	&31	&70	&\multirow{2}{*}{1.21}	\\
&&&max	&2.21	&251	&0	&44	&	\\
\hline
\noalign{\smallskip}
\multirow{2}{*}{\bf 0.8}
&\multirow{2}{*}{\bf 0}
&\multirow{2}{*}{\bf N/A}
  &min    &1.85	&255	&17	&69	&\multirow{2}{*}{0.90}	\\
&&&max    &0.95	&259	&23	&75	&	\\
\hline
\noalign{\smallskip}
\multirow{4}{*}{0.8}
&\multirow{4}{*}{50}
&\multirow{2}{*}{Prograde}
&min	&2.21	&246	&2	&54	&\multirow{2}{*}{0.97}	\\
&&&max	&1.09	&255	&18	&70	&	\\
&&\multirow{2}{*}{Retrograde}
&min	&3.22	&262	&3	&44	&\multirow{2}{*}{1.06}	\\
&&&max	&1.33	&260	&17	&66	&	\\
\hline
\noalign{\smallskip}
\multirow{2}{*}{0.9}
&\multirow{2}{*}{0}
&\multirow{2}{*}{N/A}
  &min    &2.51	&247	&7	&61	&\multirow{2}{*}{0.97}	\\
&&&max    &1.28	&253	&16	&69	&	\\
\hline
\noalign{\smallskip}
\multirow{2}{*}{1.0}
&\multirow{2}{*}{0}
&\multirow{2}{*}{N/A}
  &min    &3.08	&242	&0	&54	&\multirow{2}{*}{1.04}	\\
&&&max    &1.67	&245	&7	&63	&	\\
\hline 
\end{tabular}
\end{center}
\end{table*}

\subsection{Roughness and thermal inertia}

The shape model did not change between the IRS and MIPS observations, so the scaling factors $\gamma$ should be compatible with both IRS and MIPS data, i.e. the value for $\gamma$ (MIPS) must be between the two values of $\gamma$ (IRS) corresponding to the minimum and maximum cross-section, within the error bars. The solutions in bold in Table~\ref{table_irs} satisfy this criterion. They correspond to $\eta$=0.7 with $I$ in the range 0~--~100~J~K$^{-1}$~m$^{-2}$~s$^{-1/2}$, and to $\eta$=0.8 with a null thermal inertia. Strictly speaking, two additional solutions ($\eta$=0.7; $I$=200~J~K$^{-1}$~m$^{-2}$~s$^{-1/2}$) and ($\eta$=0.8; $I$=50~J~K$^{-1}$~m$^{-2}$~s$^{-1/2}$) are also possible, but marginally compatible with the data since they imply that we observed exactly at the maximum cross-section, moreover at the one-sigma lower limit for $\gamma$ (IRS); we therefore rejected them. For other values of $\eta$ and $I$, the difference between the scaling factor $\gamma$~(IRS) and $\gamma$~(MIPS) is too large, or the fit is unrealistic (e.g. $\mathcal{A}_{\rm SL}$=0 with no coma in the SL mode); we did not list all these solutions in Table~\ref{table_irs}.

The value of 0.7~--~0.8 for the beaming factor is low, probably indicating a high surface roughness. The nucleus thermal inertia is in the range of 0~--~100~J~K$^{-1}$~m$^{-2}$~s$^{-1/2}$, which is compatible with thermal inertia values derived for other comets, e.g., $<$45~J~K$^{-1}$~m$^{-2}$~s$^{-1/2}$ \citep{Groussin2013} and $<$200~J~K$^{-1}$~m$^{-2}$~s$^{-1/2}$ \citep{Davidsson2013} for comet 9P/Tempel~1, $<$200~J~K$^{-1}$~m$^{-2}$~s$^{-1/2}$ for comet 103P/Hartley~2 \citep{Groussin2013}, or 10~--~30~J~K$^{-1}$~m$^{-2}$~s$^{-1/2}$ \citep{Schloerb2015} and 0~--~350~J~K$^{-1}$~m$^{-2}$~s$^{-1/2}$ for comet 67P/Churyumov-Gerasimenko \citep{Marshall2018}. As suggested by \citet{Boissier2011} who derived a value $\leq$10~J~K$^{-1}$~m$^{-2}$~s$^{-1/2}$ for 8P/Tuttle from millimeter observations, the lowest values are probably more realistic. 

\subsection{Nucleus size and geometric albedo}

The scaling factor for IRS depends on the visible cross-section at the time of observation, which is unknown. As a consequence, the scaling factor derived from the MIPS thermal light curve is more robust and we chose its value to determine the size. For the valid combinations of $\eta$ and $I$ discussed above, the scaling factor lies in the range 0.83~--~0.97 or 0.90$\pm$0.07. Adding quadratically a flux calibration uncertainty of 5~\% yields $\gamma$=0.90$\pm$0.09. Applying this result to the size of the two contact spheres yields radii of 2.7$\pm$0.1~km and 1.1$\pm$0.1~km. For reference, a sphere with an ``equivalent'' radius of 2.9~km would have the same maximum cross-section.

With the Hubble Space Telescope on 10 December 2007 ($r_h$=1.26~AU, $\Delta$=0.49~AU and $\alpha$=46$^{\circ}$), \citet{Lamy2008a} derived an apparent $R$ magnitude of 15.7$\pm$0.2 corresponding to the visible light curve mean value. From this magnitude, we derived a geometric albedo of 0.042$\pm$0.008 in the $R$ band, using the above ``equivalent'' radius of 2.9~km and a linear phase correction with a phase coefficient $\beta$=0.04~mag/deg.

\subsection{Dust color temperature and $\epsilon f \rho$ quantity}

For possible solutions $(\eta,I)$, the dust temperature is well constrained to 258$\pm$10~K. This is 37~K larger than the temperature of an isothermal low-albedo dust grain in thermal equilibrium at $r$=1.6~AU from the Sun ($T\approx221$~K). This indicates that dust grains contributing to the thermal infrared flux have a typical size of $\approx$10~$\mu$m (diameter) according to \citet{Gicquel2012}, assuming porous amorphous carbon dust grains with a fractal dimension of 2.727 for the porosity model. We emphasize that it only provides a rough estimate of the grain size, due to complexity of cometary dust grains in terms of physical properties and composition \citep{Wooden2017}.

From the dust cross-section in SL and LL mode, we can derive the $\epsilon f \rho$ quantity (Section~4.3). In our case, the filling factor $f$ is equal to $\mathcal{A}_{\rm SL}$ for the SL mode or $\mathcal{A}_{\rm LL}$ for the LL mode, divided by the field of view in m$^2$. To roughly estimate the value of $\mathcal{A}_{\rm SL}$ and $\mathcal{A}_{\rm LL}$ at the time of observation, i.e. when $\gamma$ (IRS) equals $\gamma$ (MIPS) in Table~\ref{table_irs}, we interpolated linearly between the values at minimum and maximum cross-sections. For the possible solutions $(\eta,I)$, we obtain $\mathcal{A}_{\rm SL}$ in the range 23~--~28~$\times 10^6$~m$^2$ and $\mathcal{A}_{\rm LL}$ in the range 72~--~78~$\times 10^6$~m$^2$. The fields of view being $1.9\times10^{13}$~m$^2$ in the SL mode (2.6\arcsec equivalent radius) and $1.7\times10^{14}$~m$^2$ in the LL mode (7.8\arcsec equivalent radius), $\rho$ is respectively equal to 2450~km and 7440~km. Overall, we obtained $\epsilon f \rho = 314\pm31$~cm in the SL mode and $\epsilon f \rho = 305\pm13$~cm in the LL mode. Remarkably, the two values are well consistent with each other and correspond to a mean value $\epsilon f \rho = 310\pm34$~cm.

This $\epsilon f \rho$ value is similar to the value of $325\pm36$~cm derived from the MIPS images (Section~4.3), indicating that the dust production did not change significantly between the IRS observations at $r_h$=~1.6 AU pre-perihelion and the MIPS observations at $r_h$=~2.2 AU post-perihelion. The $\epsilon f \rho$ value is also larger than the $A f \rho$ values derived in the visible during the same perihelion passage (Section~1), which could indicate more dust particles of $\approx$10~$\mu$m size than of sub-micron size.
Finally, the  $\epsilon f \rho$ values of comet 8P/Tuttle are comparable to values obtained for other comets \citep{Kelley2013}.


\subsection{Water production rate}

\begin{figure} [!ht] 
\includegraphics[width=\linewidth]{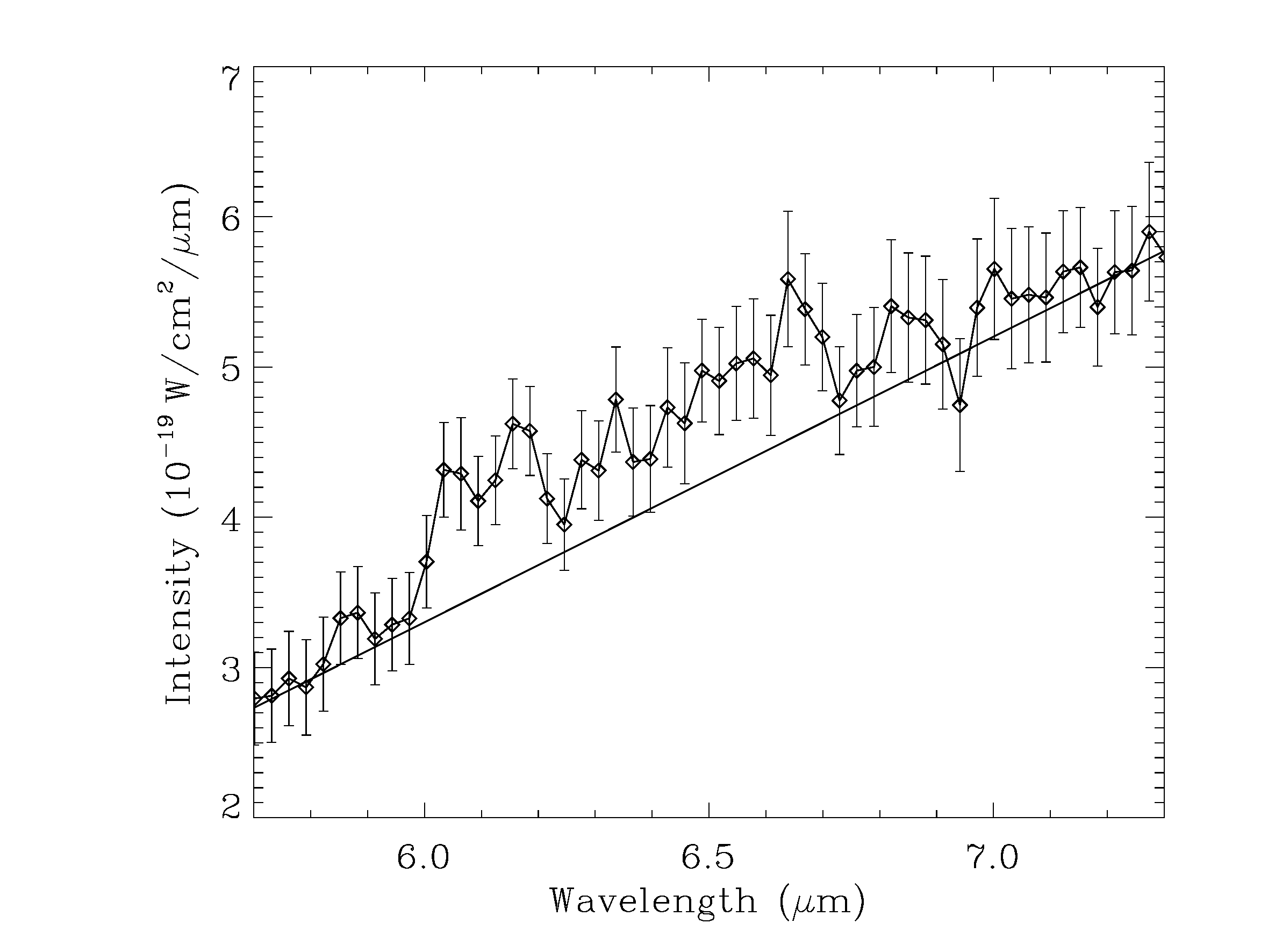}
\includegraphics[width=\linewidth]{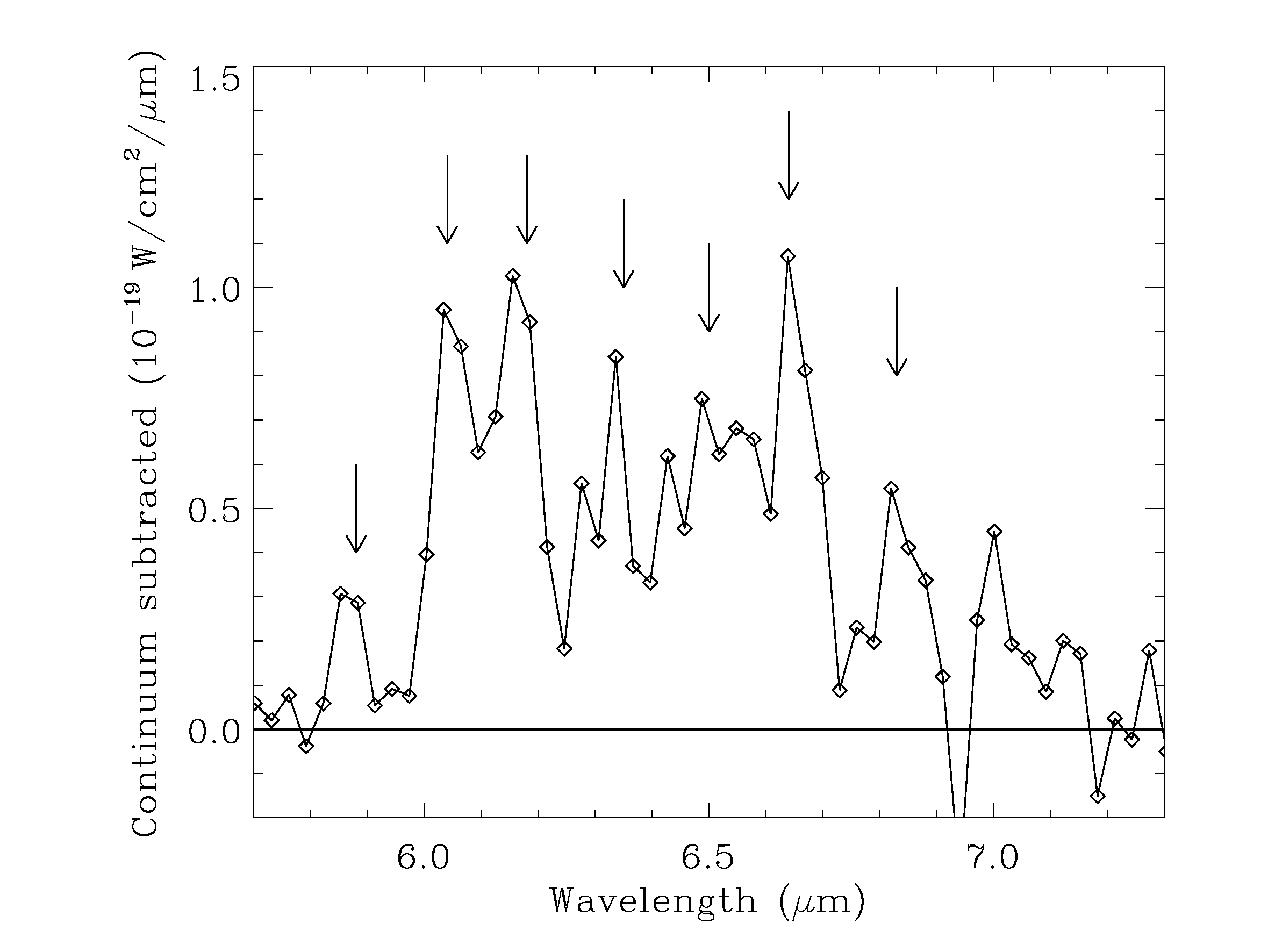}
\caption{Spectrum of the coma of 8P/Tuttle showing the $\nu_2$ water band around 6.5~$\mu$m. Upper panel: spectrum with the computed linear continuum. Lower panel: continuum subtracted spectrum, with arrows indicating the location of the strongest $\nu_2$ water emission features.}
\label{fig_water}
\end{figure}

The IRS spectrum allowed detecting the $\nu_2$ water band around 6.5~$\mu$m as illustrated in Fig.~\ref{fig_water}. The strongest water emission features are all at their expected wavelength \citep{Bockelee2009}. We modeled the continuum with a linear function, a good approximation over the small wavelength range of 5.5~--~7.3~$\mu$m. The intensity of the $\nu_2$ water band, integrated over the wavelength range 5.8~--~6.9~$\mu$m, amounts to 5.6$\pm$1.0$\times$10$^{-20}$~W~cm$^{-2}$ in an equivalent circular aperture of radius 2.85\arcsec. Using a Haser model with a g-factor of 2.41$\times$10$^{-4}$~s$^{-1}$ at $r_h$=~1~AU \citep{Bockelee2009}, and a typical gas velocity of 0.5~km~$^{-1}$ at $r_h=$~1.6~AU, we derived a water production rate of 1.1$\pm$0.2$\times$10$^{28}$~molecules~s$^{-1}$ or 340$\pm$60~kg~s$^{-1}$.

For a spherical nucleus with a radius of about 2.9~km, this implies an active fraction of $\approx$9~\%, derived from the water production rate of a spherical nucleus made of water ice only and located at the same heliocentric distance. This value is in agreement with the active fraction of 3~--~15~\% derived from the water production rates at perihelion, assuming a radius of 3~km (Section~1). The active fraction of 8P/Tuttle is comparable to that of $\approx$10~\% for 1P/Halley \citep{Keller1987}, a comet from the same dynamical family.

\subsection{Dust mineralogy}

\begin{figure} [!ht] 
  \includegraphics[width=\linewidth]{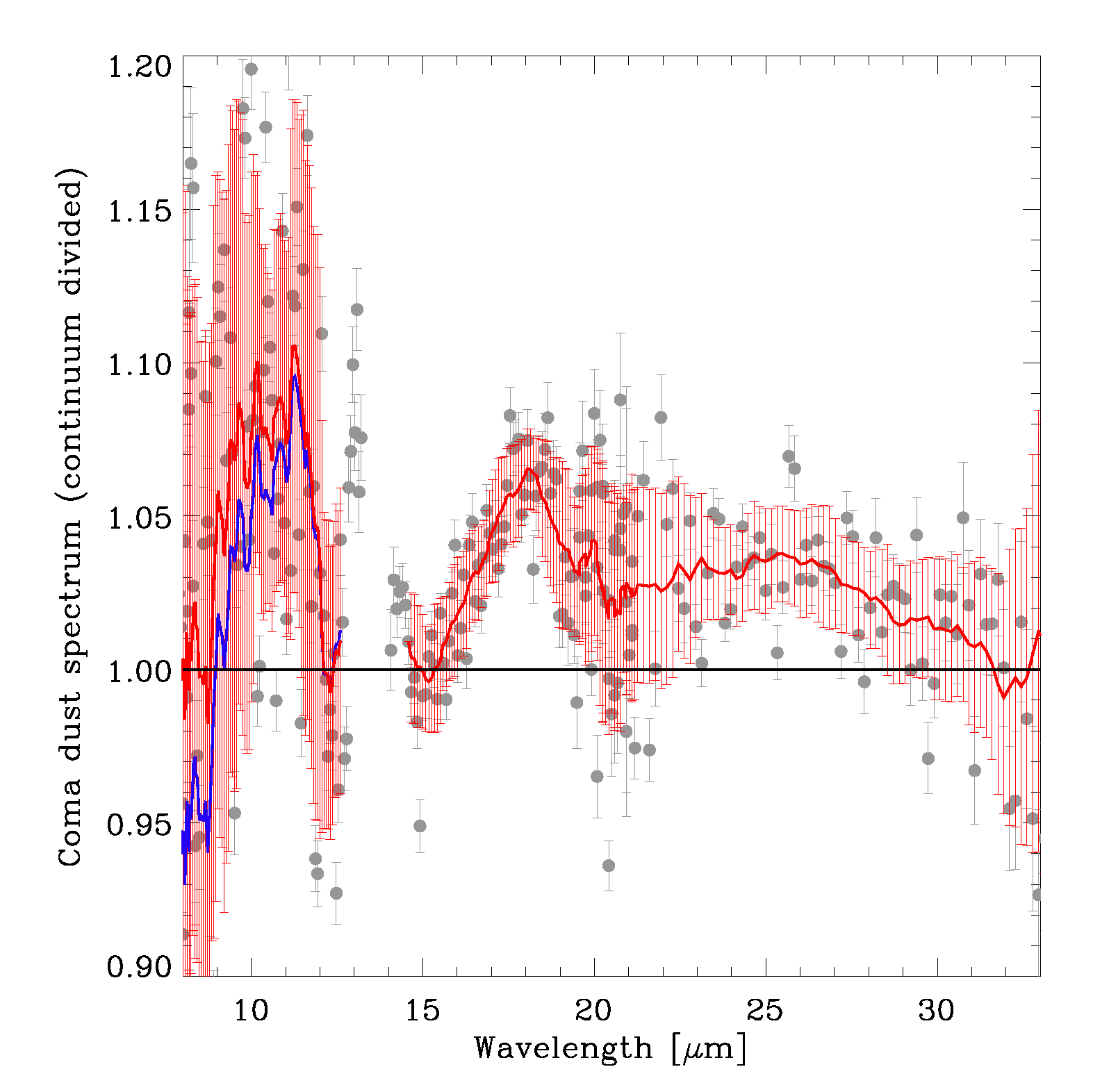}
\caption{Coma dust spectrum of comet 8P/Tuttle, continuum divided and normalized to unity at 13.0~$\mu$m. The red line corresponds to a smoothing over the data points, with a smoothing window of 15 data points. The red error bars correspond to the variance of the data points within the smoothing window. The blue line shows the spectrum in the SL mode assuming a single temperature for the dust coma, which leads to overestimating the continuum around 8~$\mu$m (see text for details).}
\label{fig_emissivity}
\end{figure}

The mid-infrared spectral domain contains features that are diagnostic of surface composition \citep{Wooden2017}. The strongest emissions are expected from amorphous silicates at $\approx$8~--~12~$\mu$m and forsterite $\approx$19.2~--~20.5~$\mu$m. Figure~\ref{fig_emissivity} shows the coma dust spectrum of comet 8P/Tuttle. This is the IRS spectrum from which we subtracted the nucleus synthetic spectrum (for the case $\eta$=0.7, $I$=50~J~K$^{-1}$~m$^{-2}$~s$^{-1/2}$, prograde rotation, and minimum cross-section) and then divided by the corresponding synthetic coma spectrum ($T_{\rm dust}$=254~K); the spectrum is normalized to unity at 13.0~$\mu$m. Since we used a single temperature for the dust coma over the full 5~--~40~$\mu$m spectral range, the continuum is overestimated around 8~$\mu$m (blue spectrum in Fig.~\ref{fig_emissivity}). To correct for this effect and obtain a better spectrum over the full wavelength range, we used a slightly lower temperature $T_{\rm dust}$=248~K in the SL range (red spectrum in Fig.~\ref{fig_emissivity}).

\begin{figure} [!ht] 
  \includegraphics[width=\linewidth]{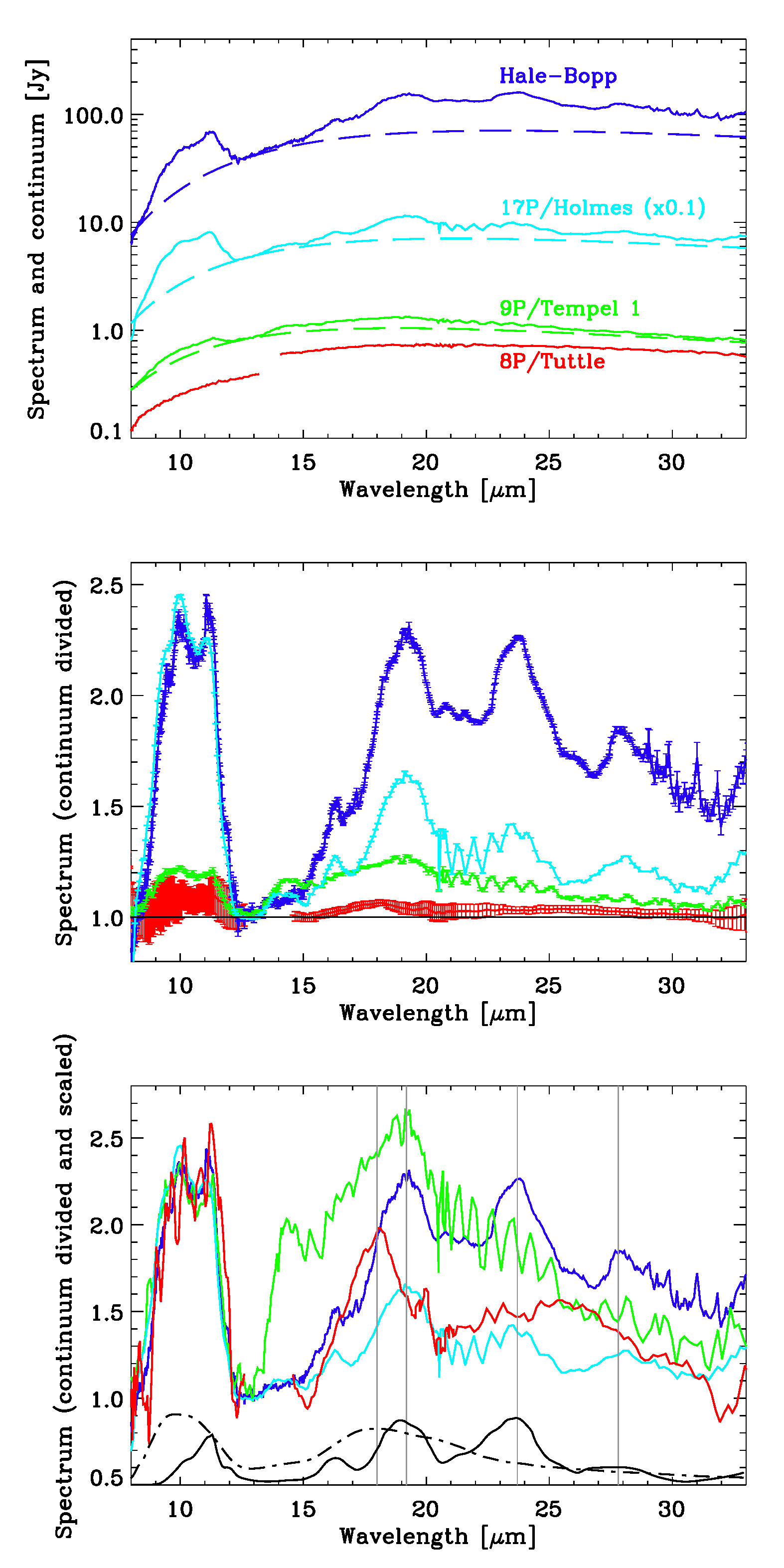}
  \caption{{\it Upper panel} - Spectra of several comets, for comparison with 8P/Tuttle. The data are from \citet{Crovisier1997} for Hale-Bopp, \citet{Lisse2006} and \citet{Kelley2009} for 9P/Tempel~1 (pre- Deep Impact), and \citet{Reach2010} for 17P/Holmes (soon after the November 2007 explosion).  The spectrum of 17P/Holmes has been divided by 10 for clarity. Dashed lines correspond to a black body continuum of 220~K for Hale-Bopp, 240~K for 17P/Holmes and 270~K for 9P/Tempel~1. The continuum for 8P/Tuttle is not shown for clarity, but identical to that of Fig.~\ref{fig_fit_irs}. {\it Middle panel} - Continuum divided spectra, using the same color code as in the upper panel. Error bars are indicated. For 8P/Tuttle, this is the same spectrum as in Fig.~\ref{fig_emissivity}. The spectrum of 8P/Tuttle has a very low contrast compared to other comets. {\it Lower panel} - Continuum divided spectra, scaled to ease the comparison, so that they all have the same value at 11~$\mu$m. Same color code as in the other panels. The black spectra on the bottom are those of crystalline forsterite (solid line) and amorphous pyroxene (dash dotted line), extracted from \citet{Reach2010} for grains with a diameter of 2~$\mu$m. The vertical gray lines highlight the position of the amorphous pyroxene emission around 18~$\mu$m observed on comet 8P/Tuttle compared with that of forsterite at 19.2~$\mu$m, 23.7~$\mu$m and 27.8~$\mu$m observed on the other comets.}
\label{fig_emissivity2}
\end{figure}

The coma dust spectrum seems to exhibit the broad amorphous silicate emission at $\approx$8~--~12~$\mu$m (1.5-$\sigma$ confidence level) already observed on several comets (Fig.~\ref{fig_emissivity2}), despite the poor SNR of the spectrum in this wavelength range. The spectrum also exhibits a second broad emission at $\approx$16~--~21~$\mu$m (5-$\sigma$ confidence level) and a large bump around 25~$\mu$m (2-$\sigma$ confidence level). The broad emissions around 10~$\mu$m and 18~$\mu$m are consistent with those of amorphous pyroxene, which exibits similar emissions at $\approx$9.0~--~11.0~$\mu$m and $\approx$16.0~--~22.0~$\mu$m \citep{Wooden2005,Reach2010} (Fig.~\ref{fig_emissivity2}, lower panel). The 18~$\mu$m emission is weak (only 6\%), which is consistent with our inferred grain size of $\approx$10~$\mu$m (diameter) since large grains reduce the contrast of emission features compared to small, micron to sub-micron size, grains. Amorphous pyroxene was also detected on comet 17P/Holmes soon after its explosion on November 2007 \citep{Reach2010}. The large bump around 25~$\mu$m is however difficult to explain in terms of mineralogical composition.

Contrary to other comets (Fig.~\ref{fig_emissivity2}), the spectrum of 8P/Tuttle lacks the forsterite features at 19.2~$\mu$m, 23.7~$\mu$m and 27.8~$\mu$m \citep{Wooden2017}. Additionally, the large bump at 25~$\mu$m is located in a region where other comets do not show any obvious emission features. These differences are puzzling and may result from the complexity of cometary dust grains in terms of physical properties (size distribution, shape, porosity, temperature) and mineralogical composition.

%
%
\section{Conclusions}

\begin{table}
  \caption[]{Properties of comet 8P/Tuttle derived from our Spitzer Space Telescope observations.}
\label{table_summary}
\begin{tabular}{ll}
\hline
\noalign{\smallskip}
{\bf Nucleus} & \\
- Shape & Two spheres in contact (binary)\\
- Size of each sphere & 2.7$\pm$0.1~km and 1.1$\pm$0.1~km (radii) \\
- Pole orientation & RA=285$\pm$12$^{\circ}$ and DEC=+20$\pm$5$^{\circ}$\\
- Geometric albedo & 0.042$\pm$0.008 ($R$-band)\\
\noalign{\smallskip}
\hline
\noalign{\smallskip}
\multicolumn{2}{l}{\bf Roughness and thermal inertia} \\
- Valid combinations       & $\eta$=0.7 and $I$=0~--~100~J~K$^{-1}$~m$^{-2}$~s$^{-1/2}$\\
                                         & or $\eta$=0.8 and $I$=0~J~K$^{-1}$~m$^{-2}$~s$^{-1/2}$\\
\noalign{\smallskip}
\hline
\noalign{\smallskip}
{\bf Production rates} & \\
- Water $Q_{\rm{H_{2}O}}$ & 1.1$\pm$0.2$\times$10$^{28}$~s$^{-1}$ (1.6~AU pre-peri.)\\
- Active fraction & $\approx$9~\% \\
- Dust $\epsilon f \rho$ & $310\pm34$~cm (1.6~AU pre-perihelion)\\
                         & $325\pm36$~cm (2.2~AU post-perihelion)\\
\noalign{\smallskip}
\hline
\noalign{\smallskip}
{\bf Dust properties} & \\
- $T_{\rm dust}$ & 258$\pm$10~K (1.6~AU pre-perihelion)\\
- Grain size & $\approx$10~$\mu$m (diameter)\\
- Composition & Amorphous pyroxene\\
\noalign{\smallskip}
\hline
\end{tabular}
\end{table}

In this paper, we have presented infrared observations of comet 8P/Tuttle performed with the MIPS and IRS instruments of the Spitzer Space Telescope. Our main results are summarized in Table~\ref{table_summary} and below.
\begin{enumerate}
\item{The HST shape model outperforms the radar shape model, providing a better qualitative and quantitative fit to the MIPS thermal light curve. This fit leads to a bilobate shape composed of two spheres in contact with radii of 2.7$\pm$0.1~km and 1.1$\pm$0.1~km and a pole orientation given by RA=285$\pm$12$^{\circ}$ and DEC=+20$\pm$5$^{\circ}$.}
\item{The $R$-band geometric albedo is 0.042$\pm$0.008.}  
\item{The thermal inertia is in the range 0~--~100~J~K$^{-1}$~m$^{-2}$~s$^{-1/2}$; within this range, we favor the lowest values.}
\item{The surface roughness is rather high with a beaming factor in the range 0.7~--~0.8.}
\item{The water production rate is estimated to be 1.1$\pm$0.2$\times$10$^{28}$~molecules~s$^{-1}$ at $r_h=$~1.6~AU pre-perihelion, which corresponds to an active fraction of $\approx$9~\%, similar to that of 1P/Halley, a comet of the same dynamical family.}
\item{The dust $\epsilon f \rho$ quantity amounts to $310\pm34$~cm at $r_h=$~1.6~AU pre-perihelion, and to $325\pm36$~cm at $r_h=$~2.2~AU post-perihelion.}
\item{The dust grain temperature is estimated to 258$\pm$10~K, which is 37~K larger than the thermal equilibrium temperature at $r_h=$~1.6~AU. This indicates that the dust grains contributing to the thermal infrared flux have a typical size of $\approx$10~$\mu$m.}
\item{The dust spectrum exhibits broad emissions around 10~$\mu$m (1.5-$\sigma$ confidence level) and 18~$\mu$m (5-$\sigma$ confidence level) that we attribute to amorphous pyroxene.}
\end{enumerate}

Despite being a Nearly Isotropic Comet, the above results do not indicate that comet 8P/Tuttle has intrinsically different physical properties than Ecliptic Comets. The size, albedo, thermal properties, water and dust production rate are not unusual compared with ECs \citep{Lamy2004,Bockelee2004}. Such similarities between NIC and EC  comets have already been observed in the past, suggesting that they ``formed in largely overlapping regions where the giant planets are today'' \citep{AHearn2012}.

The discrepancy between the HST and radar shape models comes from their different shapes {\it and} pole orientations. The RA values of the two pole directions agree within 1$\sigma$, but the DEC values are currently not compatible even at the 3$\sigma$ level. Since the HST solution (shape model + pole orientation) provides a better fit to the MIPS thermal light curve {\it and} to the HST visible light curve \citep{Lamy2008a}, it would be interesting to reanalyse the radar observations of \cite{Harmon2010} with this solution. A shape model and a pole solution that would be consistent with the radar, HST and SST observations all together may then be found.

The radius of the nucleus is more than 2.5 times smaller than expected before the perihelion passage (7.3~km, Section~1), which is quite surprising. We note that a 7.3~km radius is ruled out by our IRS and MIPS observations in any case, since the infrared flux of such a large body would exceed the observed one, even assuming no coma. The most likely explanation is that, although noted as inactive at the time of observation with a stellar profile, the contributions of the coma and of the dust tail to the central pixel were not negligible during earlier ground-based observations.

Finally, our observations are consistent with the bilobate shape of the nucleus of comet 8P/Tuttle. As noted in Section~1, this shape is likely common among comets since it was found for four out of the six comets for which we have spatially resolved images. This is also the case of the transneptunian object 2014 MU$_{69}$ (``Ultima Thule'') observed by the New Horizon spacecraft \citep{Stern2019}. This binary configuration has some implications for the formation and evolution of 8P/Tuttle. A contact binary could result from (i) the accretion at low velocity of two primordial objects \citep{Massironi2015,Davidsson2016}, (ii) the disruption of a monolithic object due to excessive spin-up resulting from non-gravitational forces or YORP\footnote{Yarkovsky-O'Keefe-Radzievskii-Paddack} effect followed by a reaccretion \citep{Bohnhardt2004,Cuk2007,Hirabayashi2016}, or (iii) the catastrophic disruption of a monolithic object by a collision followed by a reaccretion \citep{Jutzi2017,Schwartz2018}. On the one hand, with a low thermal inertia compared with NEAs, the YORP effect is low for comets, in particular for NIC which have elongated orbit and spend most of their time far from the Sun, and may not be sufficient to increase the spin rate of the nucleus to the point where centrifugal exceed gravitational forces. On the other hand, comet 8P/Tuttle has been on a very stable orbit for centuries and it is likely an evolved comet as suggested by its low activity, so that it could have been much more active in the past. Since for cometary nuclei, the primary cause for spin-up is torques caused by outgassing, it is possible that 8P/Tuttle formed as a monolithic body, and became a contact binary after its injection in the inner Solar System, due to excessive spin-up resulting from non-gravitational forces. This scenario has been proposed for comet 67P/Churyumov-Gerasimenko by \citet{Hirabayashi2016}. Alternatively, if the binary nature of comet 8P/Tuttle is the result of a primordial accretion or a catastrophic collision in the early Solar Sytem, it could have persisted until now. Similar examples are offered by some binary asteroids that can be stable over the age of the Solar System \citep{Chauvineau1991}, or as proposed by \citet{Davidsson2016} for comet 67P/Churyumov-Gerasimenko. For comet 8P/Tuttle, it is however not possible to disentangle the solution of a binary nucleus formed in the first billion year of our solar system \citep[e.g.,][]{Matonti2019} from a more recent origin following its injection in the inner solar system \citep[e.g.,][]{Hirabayashi2016}.

\begin{acknowledgements}
This work is based on observations made with the Spitzer Space Telescope, which is operated by the Jet Propulsion Laboratory, California Institute of Technology under a contract with NASA. We thank the SST ground system personnel for their prompt and efficient scheduling of the observations. We are grateful to G. Faury (IRAP, Toulouse, France) for his early contribution to the rotational analysis. I. Toth was supported by a grant from CNES (Centre National d'Etudes Spatiales) for his work at Laboratoire d'Astrophysique de Marseille. H.~Weaver gratefully acknowledges support provided by NASA through grant number GO-11226 from the Space Telescope Science Institute, which is operated by the Association of Universities for Research in Astronomy under NASA contract NAS5-26555. This research made use of Tiny~Tim/Spitzer, developed by John Krist for the Spitzer Science Center. The Center is managed by the California Institute of Technology under a contract with NASA. Finally, we thank D.~Wooden for reviewing this paper and for her constructive reports.
\end{acknowledgements}

\bibliographystyle{aa}
\bibliography{References_Tuttle}

\begin{thebibliography}{67}
\expandafter\ifx\csname natexlab\endcsname\relax\def\natexlab#1{#1}\fi

\bibitem[{{A'Hearn} {et~al.}(2011){A'Hearn}, {Belton}, {Delamere}, {Feaga},
  {Hampton}, {Kissel}, {Klaasen}, {McFadden}, {Meech}, {Melosh}, {Schultz},
  {Sunshine}, {Thomas}, {Veverka}, {Wellnitz}, {Yeomans}, {Besse}, {Bodewits},
  {Bowling}, {Carcich}, {Collins}, {Farnham}, {Groussin}, {Hermalyn}, {Kelley},
  {Kelley}, {Li}, {Lindler}, {Lisse}, {McLaughlin}, {Merlin}, {Protopapa},
  {Richardson}, \& {Williams}}]{AHearn2011}
{A'Hearn}, M.~F., {Belton}, M.~J.~S., {Delamere}, W.~A., {et~al.} 2011,
  Science, 332, 1396

\bibitem[{{A'Hearn} {et~al.}(2012){A'Hearn}, {Feaga}, {Keller}, {Kawakita},
  {Hampton}, {Kissel}, {Klaasen}, {McFadden}, {Meech}, {Schultz}, {Sunshine},
  {Thomas}, {Veverka}, {Yeomans}, {Besse}, {Bodewits}, {Farnham}, {Groussin},
  {Kelley}, {Lisse}, {Merlin}, {Protopapa}, \& {Wellnitz}}]{AHearn2012}
{A'Hearn}, M.~F., {Feaga}, L.~M., {Keller}, H.~U., {et~al.} 2012, \apj, 758, 29

\bibitem[{{A'Hearn} {et~al.}(1995){A'Hearn}, {Millis}, {Schleicher}, {Osip}, \&
  {Birch}}]{AHearn1995}
{A'Hearn}, M.~F., {Millis}, R.~C., {Schleicher}, D.~O., {Osip}, D.~J., \&
  {Birch}, P.~V. 1995, \icarus, 118, 223

\bibitem[{{A'Hearn} {et~al.}(1984){A'Hearn}, {Schleicher}, {Millis}, {Feldman},
  \& {Thompson}}]{AHearn1984}
{A'Hearn}, M.~F., {Schleicher}, D.~G., {Millis}, R.~L., {Feldman}, P.~D., \&
  {Thompson}, D.~T. 1984, \aj, 89, 579

\bibitem[{{Barber} {et~al.}(2009){Barber}, {Miller}, {Dello Russo}, {Mumma},
  {Tennyson}, \& {Guio}}]{Barber2009}
{Barber}, R.~J., {Miller}, S., {Dello Russo}, N., {et~al.} 2009, \mnras, 398,
  1593

\bibitem[{{Biver} {et~al.}(2008){Biver}, {Lis}, {Fray}, {Bockel{\'e}e-Morvan},
  {Crovisier}, {Boissier}, {Colom}, {Dello-Russo}, {Moreno}, {Vervack}, \&
  {Weaver}}]{Biver2008}
{Biver}, N., {Lis}, D.~C., {Fray}, N., {et~al.} 2008, in LPI Contributions,
  Vol. 1405, Asteroids, Comets, Meteors 2008, 8151

\bibitem[{{Bockel{\'e}e-Morvan} {et~al.}(2004){Bockel{\'e}e-Morvan},
  {Crovisier}, {Mumma}, \& {Weaver}}]{Bockelee2004}
{Bockel{\'e}e-Morvan}, D., {Crovisier}, J., {Mumma}, M.~J., \& {Weaver}, H.~A.
  2004, {The composition of cometary volatiles}, ed. M.~C. {Festou}, H.~U.
  {Keller}, \& H.~A. {Weaver}, 391--423

\bibitem[{{Bockel{\'e}e-Morvan} {et~al.}(2008){Bockel{\'e}e-Morvan}, {Dello
  Russo}, {Jehin}, {Manfroid}, {Smette}, {Cochran}, {Hutsem{\'e}kers},
  {Kawakita}, {Kobayashi}, {Schulz}, {Weiler}, {Zucconi}, {Arpigny}, {Biver},
  {Crovisier}, {Magain}, {Rauer}, {Sana}, {Vervack}, \&
  {Weaver}}]{Bockelee2008}
{Bockel{\'e}e-Morvan}, D., {Dello Russo}, N., {Jehin}, E., {et~al.} 2008, in
  LPI Contributions, Vol. 1405, Asteroids, Comets, Meteors 2008, 8190

\bibitem[{{Bockel{\'e}e-Morvan} {et~al.}(2009){Bockel{\'e}e-Morvan},
  {Woodward}, {Kelley}, \& {Wooden}}]{Bockelee2009}
{Bockel{\'e}e-Morvan}, D., {Woodward}, C.~E., {Kelley}, M.~S., \& {Wooden},
  D.~H. 2009, \apj, 696, 1075

\bibitem[{{Boehnhardt}(2004)}]{Bohnhardt2004}
{Boehnhardt}, H. 2004, {Split comets}, ed. M.~C. {Festou}, H.~U. {Keller}, \&
  H.~A. {Weaver}, 301--316

\bibitem[{{B{\"o}hnhardt} {et~al.}(2008){B{\"o}hnhardt}, {Mumma}, {Villanueva},
  {DiSanti}, {Bonev}, {Lippi}, \& {K{\"a}ufl}}]{Bohnhardt2008}
{B{\"o}hnhardt}, H., {Mumma}, M.~J., {Villanueva}, G.~L., {et~al.} 2008, \apjl,
  683, L71

\bibitem[{{Boissier} {et~al.}(2011){Boissier}, {Groussin}, {Jorda}, {Lamy},
  {Bockel{\'e}e-Morvan}, {Crovisier}, {Biver}, {Colom}, {Lellouch}, \&
  {Moreno}}]{Boissier2011}
{Boissier}, J., {Groussin}, O., {Jorda}, L., {et~al.} 2011, \aap, 528, A54

\bibitem[{{Bonev} {et~al.}(2008){Bonev}, {Mumma}, {Radeva}, {DiSanti}, {Gibb},
  \& {Villanueva}}]{Bonev2008}
{Bonev}, B.~P., {Mumma}, M.~J., {Radeva}, Y.~L., {et~al.} 2008, \apjl, 680, L61

\bibitem[{{Buratti} {et~al.}(2004){Buratti}, {Hicks}, {Soderblom}, {Britt},
  {Oberst}, \& {Hillier}}]{Buratti2004}
{Buratti}, B.~J., {Hicks}, M.~D., {Soderblom}, L.~A., {et~al.} 2004, \icarus,
  167, 16

\bibitem[{{Chauvineau} {et~al.}(1991){Chauvineau}, {Mignard}, \&
  {Farinella}}]{Chauvineau1991}
{Chauvineau}, B., {Mignard}, F., \& {Farinella}, P. 1991, \icarus, 94, 299

\bibitem[{{Crovisier} {et~al.}(1997){Crovisier}, {Leech}, {Bockelee-Morvan},
  {Brooke}, {Hanner}, {Altieri}, {Keller}, \& {Lellouch}}]{Crovisier1997}
{Crovisier}, J., {Leech}, K., {Bockelee-Morvan}, D., {et~al.} 1997, Science,
  275, 1904

\bibitem[{{{\'C}uk}(2007)}]{Cuk2007}
{{\'C}uk}, M. 2007, \apjl, 659, L57

\bibitem[{{Davidsson} {et~al.}(2013){Davidsson}, {Guti{\'e}rrez}, {Groussin},
  {A'Hearn}, {Farnham}, {Feaga}, {Kelley}, {Klaasen}, {Merlin}, {Protopapa},
  {Rickman}, {Sunshine}, \& {Thomas}}]{Davidsson2013}
{Davidsson}, B.~J.~R., {Guti{\'e}rrez}, P.~J., {Groussin}, O., {et~al.} 2013,
  \icarus, 224, 154

\bibitem[{{Davidsson} {et~al.}(2016){Davidsson}, {Sierks}, {G{\"u}ttler},
  {Marzari}, {Pajola}, {Rickman}, {A'Hearn}, {Auger}, {El-Maarry}, {Fornasier},
  {Guti{\'e}rrez}, {Keller}, {Massironi}, {Snodgrass}, {Vincent}, {Barbieri},
  {Lamy}, {Rodrigo}, {Koschny}, {Barucci}, {Bertaux}, {Bertini}, {Cremonese},
  {Da Deppo}, {Debei}, {De Cecco}, {Feller}, {Fulle}, {Groussin}, {Hviid},
  {H{\"o}fner}, {Ip}, {Jorda}, {Knollenberg}, {Kovacs}, {Kramm}, {K{\"u}hrt},
  {K{\"u}ppers}, {La Forgia}, {Lara}, {Lazzarin}, {Lopez Moreno},
  {Moissl-Fraund}, {Mottola}, {Naletto}, {Oklay}, {Thomas}, \&
  {Tubiana}}]{Davidsson2016}
{Davidsson}, B.~J.~R., {Sierks}, H., {G{\"u}ttler}, C., {et~al.} 2016, \aap,
  592, A63

\bibitem[{{Fern{\'a}ndez} {et~al.}(2013){Fern{\'a}ndez}, {Kelley}, {Lamy},
  {Toth}, {Groussin}, {Lisse}, {A'Hearn}, {Bauer}, {Campins}, {Fitzsimmons},
  {Licandro}, {Lowry}, {Meech}, {Pittichov{\'a}}, {Reach}, {Snodgrass}, \&
  {Weaver}}]{Fernandez2013}
{Fern{\'a}ndez}, Y.~R., {Kelley}, M.~S., {Lamy}, P.~L., {et~al.} 2013, \icarus,
  226, 1138

\bibitem[{{Gicquel} {et~al.}(2012){Gicquel}, {Bockel{\'e}e-Morvan}, {Zakharov},
  {Kelley}, {Woodward}, \& {Wooden}}]{Gicquel2012}
{Gicquel}, A., {Bockel{\'e}e-Morvan}, D., {Zakharov}, V.~V., {et~al.} 2012,
  \aap, 542, A119

\bibitem[{{Groussin} {et~al.}(2007){Groussin}, {A'Hearn}, {Li}, {Thomas},
  {Sunshine}, {Lisse}, {Meech}, {Farnham}, {Feaga}, \&
  {Delamere}}]{Groussin2007}
{Groussin}, O., {A'Hearn}, M.~F., {Li}, J.-Y., {et~al.} 2007, \icarus, 187, 16

\bibitem[{{Groussin} {et~al.}(2019){Groussin}, {Attree}, {Brouet}, {Ciarletti},
  {Davidsson}, {Filacchione}, {Fischer}, {Gundlach}, {Knapmeyer},
  {Knollenberg}, {Kokotanekova}, {K{\"u}hrt}, {Leyrat}, {Marshall}, {Pelivan},
  {Skorov}, {Snodgrass}, {Spohn}, \& {Tosi}}]{Groussin2019}
{Groussin}, O., {Attree}, N., {Brouet}, Y., {et~al.} 2019, \ssr, 215, 29

\bibitem[{{Groussin} {et~al.}(2004){Groussin}, {Lamy}, \&
  {Jorda}}]{Groussin2004}
{Groussin}, O., {Lamy}, P., \& {Jorda}, L. 2004, \aap, 413, 1163

\bibitem[{{Groussin} {et~al.}(2013){Groussin}, {Sunshine}, {Feaga}, {Jorda},
  {Thomas}, {Li}, {A'Hearn}, {Belton}, {Besse}, {Carcich}, {Farnham},
  {Hampton}, {Klaasen}, {Lisse}, {Merlin}, \& {Protopapa}}]{Groussin2013}
{Groussin}, O., {Sunshine}, J.~M., {Feaga}, L.~M., {et~al.} 2013, \icarus, 222,
  580

\bibitem[{{Harmon} {et~al.}(2010){Harmon}, {Nolan}, {Giorgini}, \&
  {Howell}}]{Harmon2010}
{Harmon}, J.~K., {Nolan}, M.~C., {Giorgini}, J.~D., \& {Howell}, E.~S. 2010,
  \icarus, 207, 499

\bibitem[{{Harris}(1998)}]{Harris1998}
{Harris}, A.~W. 1998, Icarus, 131, 291

\bibitem[{{Hirabayashi} {et~al.}(2016){Hirabayashi}, {Scheeres}, {Chesley},
  {Marchi}, {McMahon}, {Steckloff}, {Mottola}, {Naidu}, \&
  {Bowling}}]{Hirabayashi2016}
{Hirabayashi}, M., {Scheeres}, D.~J., {Chesley}, S.~R., {et~al.} 2016, \nat,
  534, 352

\bibitem[{{Houck} {et~al.}(2004){Houck}, {Roellig}, {van Cleve}, {Forrest},
  {Herter}, {Lawrence}, {Matthews}, {Reitsema}, {Soifer}, {Watson}, {Weedman},
  {Huisjen}, {Troeltzsch}, {Barry}, {Bernard-Salas}, {Blacken}, {Brandl},
  {Charmandaris}, {Devost}, {Gull}, {Hall}, {Henderson}, {Higdon}, {Pirger},
  {Schoenwald}, {Sloan}, {Uchida}, {Appleton}, {Armus}, {Burgdorf},
  {Fajardo-Acosta}, {Grillmair}, {Ingalls}, {Morris}, \& {Teplitz}}]{Houck2004}
{Houck}, J.~R., {Roellig}, T.~L., {van Cleve}, J., {et~al.} 2004, \apjs, 154,
  18

\bibitem[{{Hui} \& {Li}(2018)}]{Hui2018}
{Hui}, M.-T. \& {Li}, J.-Y. 2018, \pasp, 130, 104501

\bibitem[{{Jehin} {et~al.}(2009){Jehin}, {Bockel{\'e}e-Morvan}, {Dello Russo},
  {Manfroid}, {Hutsem{\'e}kers}, {Kawakita}, {Kobayashi}, {Schulz}, {Smette},
  {St{\"u}we}, {Weiler}, {Arpigny}, {Biver}, {Cochran}, {Crovisier}, {Magain},
  {Rauer}, {Sana}, {Vervack}, {Weaver}, \& {Zucconi}}]{Jehin2009}
{Jehin}, E., {Bockel{\'e}e-Morvan}, D., {Dello Russo}, N., {et~al.} 2009, Earth
  Moon and Planets, 105, 343

\bibitem[{{Jutzi} \& {Benz}(2017)}]{Jutzi2017}
{Jutzi}, M. \& {Benz}, W. 2017, \aap, 597, A62

\bibitem[{{Keller} {et~al.}(1987){Keller}, {Delamere}, {Reitsema}, {Huebner},
  \& {Schmidt}}]{Keller1987}
{Keller}, H.~U., {Delamere}, W.~A., {Reitsema}, H.~J., {Huebner}, W.~F., \&
  {Schmidt}, H.~U. 1987, \aap, 187, 807

\bibitem[{{Kelley} {et~al.}(2013){Kelley}, {Fern{\'a}ndez}, {Licandro},
  {Lisse}, {Reach}, {A'Hearn}, {Bauer}, {Campins}, {Fitzsimmons}, {Groussin},
  {Lamy}, {Lowry}, {Meech}, {Pittichov{\'a}}, {Snodgrass}, {Toth}, \&
  {Weaver}}]{Kelley2013}
{Kelley}, M.~S., {Fern{\'a}ndez}, Y.~R., {Licandro}, J., {et~al.} 2013,
  \icarus, 225, 475

\bibitem[{{Kelley} \& {Wooden}(2009)}]{Kelley2009}
{Kelley}, M.~S. \& {Wooden}, D.~H. 2009, \planss, 57, 1133

\bibitem[{{Kobayashi} {et~al.}(2010){Kobayashi}, {Bockel{\'e}e-Morvan},
  {Kawakita}, {Dello Russo}, {Jehin}, {Manfroid}, {Smette}, {Hutsem{\'e}kers},
  {St{\"u}we}, {Weiler}, {Arpigny}, {Biver}, {Cochran}, {Crovisier}, {Magain},
  {Sana}, {Schulz}, {Vervack}, {Weaver}, \& {Zucconi}}]{Kobayashi2010}
{Kobayashi}, H., {Bockel{\'e}e-Morvan}, D., {Kawakita}, H., {et~al.} 2010,
  \aap, 509, A80

\bibitem[{{Lagerros}(1998)}]{Lagerros1998}
{Lagerros}, J.~S.~V. 1998, \aap, 332, 1123

\bibitem[{{Lamy} {et~al.}(2008{\natexlab{a}}){Lamy}, {Jorda}, {Fornasier},
  {Groussin}, {Barucci}, {Carvano}, {Dotto}, {Fulchignoni}, \&
  {Toth}}]{Lamy2008c}
{Lamy}, P.~L., {Jorda}, L., {Fornasier}, S., {et~al.} 2008{\natexlab{a}}, \aap,
  487, 1187

\bibitem[{{Lamy} {et~al.}(2004){Lamy}, {Toth}, {Fernandez}, \&
  {Weaver}}]{Lamy2004}
{Lamy}, P.~L., {Toth}, I., {Fernandez}, Y.~R., \& {Weaver}, H.~A. 2004, {The
  sizes, shapes, albedos, and colors of cometary nuclei}, ed. M.~C. {Festou},
  H.~U. {Keller}, \& H.~A. {Weaver}, 223--264

\bibitem[{{Lamy} {et~al.}(2008{\natexlab{b}}){Lamy}, {Toth}, {Groussin},
  {Jorda}, {Kelley}, \& {Stansberry}}]{Lamy2008b}
{Lamy}, P.~L., {Toth}, I., {Groussin}, O., {et~al.} 2008{\natexlab{b}}, \aap,
  489, 777

\bibitem[{{Lamy} {et~al.}(2008{\natexlab{c}}){Lamy}, {Toth}, {Jorda}, {Weaver},
  {Groussin}, \& {A'Hearn}}]{Lamy2008a}
{Lamy}, P.~L., {Toth}, I., {Jorda}, L., {et~al.} 2008{\natexlab{c}}, in
  Bulletin of the American Astronomical Society, Vol.~40, AAS/Division for
  Planetary Sciences Meeting Abstracts \#40, 393

\bibitem[{{Lebofsky} \& {Spencer}(1989)}]{Lebofsky1989}
{Lebofsky}, L.~A. \& {Spencer}, J.~R. 1989, in Asteroids II, ed. R.~P.
  {Binzel}, T.~{Gehrels}, \& M.~S. {Matthews}, 128--147

\bibitem[{{Lebofsky} {et~al.}(1986){Lebofsky}, {Sykes}, {Tedesco}, {Veeder},
  {Matson}, {Brown}, {Gradie}, {Feierberg}, \& {Rudy}}]{Lebofsky1986}
{Lebofsky}, L.~A., {Sykes}, M.~V., {Tedesco}, E.~F., {et~al.} 1986, \icarus,
  68, 239

\bibitem[{{Levison}(1991)}]{Levison1991}
{Levison}, H.~F. 1991, \aj, 102, 787

\bibitem[{{Levison} {et~al.}(2001){Levison}, {Dones}, \&
  {Duncan}}]{Levison2001}
{Levison}, H.~F., {Dones}, L., \& {Duncan}, M.~J. 2001, \aj, 121, 2253

\bibitem[{{Levison} \& {Duncan}(1994)}]{Levison1994}
{Levison}, H.~F. \& {Duncan}, M.~J. 1994, \icarus, 108, 18

\bibitem[{{Levison} \& {Duncan}(1997)}]{Levison1997}
{Levison}, H.~F. \& {Duncan}, M.~J. 1997, \icarus, 127, 13

\bibitem[{{Licandro} {et~al.}(2000){Licandro}, {Tancredi}, {Lindgren},
  {Rickman}, \& {Hutton}}]{Licandro2000}
{Licandro}, J., {Tancredi}, G., {Lindgren}, M., {Rickman}, H., \& {Hutton},
  R.~G. 2000, \icarus, 147, 161

\bibitem[{{Lippi} {et~al.}(2008){Lippi}, {Mumma}, {Villanueva}, {di Santi},
  {Bonev}, \& {Boehnhardt}}]{Lippi2008}
{Lippi}, M., {Mumma}, M.~J., {Villanueva}, G.~L., {et~al.} 2008, in LPI
  Contributions, Vol. 1405, Asteroids, Comets, Meteors 2008, 8197

\bibitem[{{Lisse} {et~al.}(2005){Lisse}, {A'Hearn}, {Groussin},
  {Fern{\'a}ndez}, {Belton}, {van Cleve}, {Charmandaris}, {Meech}, \&
  {McGleam}}]{Lisse2005}
{Lisse}, C.~M., {A'Hearn}, M.~F., {Groussin}, O., {et~al.} 2005, \apjl, 625,
  L139

\bibitem[{{Lisse} {et~al.}(2006){Lisse}, {VanCleve}, {Adams}, {A'Hearn},
  {Fern{\'a}ndez}, {Farnham}, {Armus}, {Grillmair}, {Ingalls}, {Belton},
  {Groussin}, {McFadden}, {Meech}, {Schultz}, {Clark}, {Feaga}, \&
  {Sunshine}}]{Lisse2006}
{Lisse}, C.~M., {VanCleve}, J., {Adams}, A.~C., {et~al.} 2006, Science, 313,
  635

\bibitem[{{Lovell} \& {Howell}(2008)}]{Lovell2008}
{Lovell}, A.~J. \& {Howell}, E.~S. 2008, in LPI Contributions, Vol. 1405,
  Asteroids, Comets, Meteors 2008, 8379

\bibitem[{{Marshall} {et~al.}(2018){Marshall}, {Groussin}, {Vincent}, {Brouet},
  {Kappel}, {Arnold}, {Capria}, {Filacchione}, {Hartogh}, {Hofstadter}, {Ip},
  {Jorda}, {K{\"u}hrt}, {Lellouch}, {Mottola}, {Rezac}, {Rodrigo}, {Rodionov},
  {Schloerb}, \& {Thomas}}]{Marshall2018}
{Marshall}, D., {Groussin}, O., {Vincent}, J.-B., {et~al.} 2018, \aap, 616,
  A122

\bibitem[{{Massironi} {et~al.}(2015){Massironi}, {Simioni}, {Marzari},
  {Cremonese}, {Giacomini}, {Pajola}, {Jorda}, {Naletto}, {Lowry}, {El-Maarry},
  {Preusker}, {Scholten}, {Sierks}, {Barbieri}, {Lamy}, {Rodrigo}, {Koschny},
  {Rickman}, {Keller}, {A'Hearn}, {Agarwal}, {Auger}, {Barucci}, {Bertaux},
  {Bertini}, {Besse}, {Bodewits}, {Capanna}, {da Deppo}, {Davidsson}, {Debei},
  {de Cecco}, {Ferri}, {Fornasier}, {Fulle}, {Gaskell}, {Groussin},
  {Guti{\'e}rrez}, {G{\"u}ttler}, {Hviid}, {Ip}, {Knollenberg}, {Kovacs},
  {Kramm}, {K{\"u}hrt}, {K{\"u}ppers}, {La Forgia}, {Lara}, {Lazzarin}, {Lin},
  {Lopez Moreno}, {Magrin}, {Michalik}, {Mottola}, {Oklay}, {Pommerol},
  {Thomas}, {Tubiana}, \& {Vincent}}]{Massironi2015}
{Massironi}, M., {Simioni}, E., {Marzari}, F., {et~al.} 2015, \nat, 526, 402

\bibitem[{{Matonti} {et~al.}(2019){Matonti}, {Attree}, {Groussin}, {Jorda},
  {Viseur}, {Nebouy}, {Auger}, \& {Lamy}}]{Matonti2019}
{Matonti}, C., {Attree}, N., {Groussin}, O., {et~al.} 2019, Nature Geoscience,
  12, 157

\bibitem[{{Reach} {et~al.}(2010){Reach}, {Vaubaillon}, {Lisse}, {Holloway}, \&
  {Rho}}]{Reach2010}
{Reach}, W.~T., {Vaubaillon}, J., {Lisse}, C.~M., {Holloway}, M., \& {Rho}, J.
  2010, \icarus, 208, 276

\bibitem[{{Rieke} {et~al.}(2004){Rieke}, {Young}, {Engelbracht}, {Kelly},
  {Low}, {Haller}, {Beeman}, {Gordon}, {Stansberry}, {Misselt}, {Cadien},
  {Morrison}, {Rivlis}, {Latter}, {Noriega-Crespo}, {Padgett}, {Stapelfeldt},
  {Hines}, {Egami}, {Muzerolle}, {Alonso-Herrero}, {Blaylock}, {Dole}, {Hinz},
  {Le Floc'h}, {Papovich}, {P{\'e}rez-Gonz{\'a}lez}, {Smith}, {Su}, {Bennett},
  {Frayer}, {Henderson}, {Lu}, {Masci}, {Pesenson}, {Rebull}, {Rho}, {Keene},
  {Stolovy}, {Wachter}, {Wheaton}, {Werner}, \& {Richards}}]{Rieke2004}
{Rieke}, G.~H., {Young}, E.~T., {Engelbracht}, C.~W., {et~al.} 2004, \apjs,
  154, 25

\bibitem[{{Schleicher}(2007)}]{Schleicher2007}
{Schleicher}, D. 2007, \iaucirc, 8903

\bibitem[{{Schloerb} {et~al.}(2015){Schloerb}, {Keihm}, {von Allmen},
  {Choukroun}, {Lellouch}, {Leyrat}, {Beaudin}, {Biver}, {Bockel{\'e}e-Morvan},
  {Crovisier}, {Encrenaz}, {Gaskell}, {Gulkis}, {Hartogh}, {Hofstadter}, {Ip},
  {Janssen}, {Jarchow}, {Jorda}, {Keller}, {Lee}, {Rezac}, \&
  {Sierks}}]{Schloerb2015}
{Schloerb}, F.~P., {Keihm}, S., {von Allmen}, P., {et~al.} 2015, \aap, 583, A29

\bibitem[{{Schwartz} {et~al.}(2018){Schwartz}, {Michel}, {Jutzi}, {Marchi},
  {Zhang}, \& {Richardson}}]{Schwartz2018}
{Schwartz}, S.~R., {Michel}, P., {Jutzi}, M., {et~al.} 2018, Nature Astronomy,
  2, 379

\bibitem[{{Sierks} {et~al.}(2015){Sierks}, {Barbieri}, {Lamy}, {Rodrigo},
  {Koschny}, {Rickman}, {Keller}, {Agarwal}, {A'Hearn}, {Angrilli}, {Auger},
  {Barucci}, {Bertaux}, {Bertini}, {Besse}, {Bodewits}, {Capanna}, {Cremonese},
  {Da Deppo}, {Davidsson}, {Debei}, {De Cecco}, {Ferri}, {Fornasier}, {Fulle},
  {Gaskell}, {Giacomini}, {Groussin}, {Gutierrez-Marques}, {Guti{\'e}rrez},
  {G{\"u}ttler}, {Hoekzema}, {Hviid}, {Ip}, {Jorda}, {Knollenberg}, {Kovacs},
  {Kramm}, {K{\"u}hrt}, {K{\"u}ppers}, {La Forgia}, {Lara}, {Lazzarin},
  {Leyrat}, {Lopez Moreno}, {Magrin}, {Marchi}, {Marzari}, {Massironi},
  {Michalik}, {Moissl}, {Mottola}, {Naletto}, {Oklay}, {Pajola}, {Pertile},
  {Preusker}, {Sabau}, {Scholten}, {Snodgrass}, {Thomas}, {Tubiana}, {Vincent},
  {Wenzel}, {Zaccariotto}, \& {P{\"a}tzold}}]{Sierks2015}
{Sierks}, H., {Barbieri}, C., {Lamy}, P.~L., {et~al.} 2015, Science, 347,
  aaa1044

\bibitem[{{Soderblom} {et~al.}(2002){Soderblom}, {Becker}, {Bennett}, {Boice},
  {Britt}, {Brown}, {Buratti}, {Isbell}, {Giese}, {Hare}, {Hicks},
  {Howington-Kraus}, {Kirk}, {Lee}, {Nelson}, {Oberst}, {Owen}, {Rayman},
  {Sandel}, {Stern}, {Thomas}, \& {Yelle}}]{Soderblom2002}
{Soderblom}, L.~A., {Becker}, T.~L., {Bennett}, G., {et~al.} 2002, Science,
  296, 1087

\bibitem[{{Stern} {et~al.}(2019){Stern}, {Weaver}, {Spencer}, {Olkin},
  {Gladstone}, {Grundy}, {Moore}, {Cruikshank}, {Elliott}, {McKinnon}, \&
  et~al.}]{Stern2019}
{Stern}, S.~A., {Weaver}, H.~A., {Spencer}, J.~R., {et~al.} 2019, Science, 364,
  aaw9771

\bibitem[{{Weissman} {et~al.}(2008){Weissman}, {Choi}, \&
  {Lowry}}]{Weissman2008}
{Weissman}, P.~R., {Choi}, Y.~J., \& {Lowry}, S.~C. 2008, in Bulletin of the
  American Astronomical Society, Vol.~40, AAS/Division for Planetary Sciences
  Meeting Abstracts \#40, 387

\bibitem[{{Werner} {et~al.}(2004){Werner}, {Roellig}, {Low}, {Rieke}, {Rieke},
  {Hoffmann}, {Young}, {Houck}, {Brandl}, {Fazio}, {Hora}, {Gehrz}, {Helou},
  {Soifer}, {Stauffer}, {Keene}, {Eisenhardt}, {Gallagher}, {Gautier}, {Irace},
  {Lawrence}, {Simmons}, {Van Cleve}, {Jura}, {Wright}, \&
  {Cruikshank}}]{Werner2004}
{Werner}, M.~W., {Roellig}, T.~L., {Low}, F.~J., {et~al.} 2004, \apjs, 154, 1

\bibitem[{{Wooden} {et~al.}(2005){Wooden}, {Harker}, \&
  {Brearley}}]{Wooden2005}
{Wooden}, D.~H., {Harker}, D.~E., \& {Brearley}, A.~J. 2005, in Astronomical
  Society of the Pacific Conference Series, Vol. 341, Chondrites and the
  Protoplanetary Disk, ed. A.~N. {Krot}, E.~R.~D. {Scott}, \& B.~{Reipurth},
  774

\bibitem[{{Wooden} {et~al.}(2017){Wooden}, {Ishii}, \& {Zolensky}}]{Wooden2017}
{Wooden}, D.~H., {Ishii}, H.~A., \& {Zolensky}, M.~E. 2017, Philosophical
  Transactions of the Royal Society of London Series A, 375, 20160260

\end{thebibliography}

\end{document}